\documentclass[twocolumn]{aastex631}

\newcommand{\mfp}{$\lambda_{\text{mfp}}$ }
\shorttitle{Individual free paths at $z\sim6$}
\shortauthors{S.~E.~I.~Bosman}
\usepackage{bigints}

\begin{document}

\title{Constraints on the mean free path of ionising photons at $z\sim6$ using limits on individual free paths}

\correspondingauthor{Sarah E.~I.~Bosman}
\email{bosman@mpia.de}

\author[0000-0001-8582-7012]{Sarah E.~I.~Bosman}
\affiliation{Max-Planck-Institut f{\"u}r Astronomie, K{\"o}nigstuhl 17, D-69117 Heidelberg, Germany}



\begin{abstract}

The recent measurement of an ionising mean free path $\lambda_{\text{mfp}}<1$ pMpc at $z=6$ challenges our understanding of the small-scale structure of the intergalactic medium (IGM) at the end of reionisation. We introduce a new method to constrain \mfp at $z=6$ by using lower limits on the individual free paths of ionisation around quasars. Lyman-limit absorbers with a density sufficient to halt ionising photons produce strong absorption in the 6 lowest-energy Lyman transitions, in the absence of which a robust lower limit can be placed on the individual free path. Applying this method to a set of $26$ quasars at $5.5<z<6.5$, we find that $80\%$ of bright quasars ($M_{1450}<-26.5$) require individual free paths larger than $2$ pMpc. We model the relation between opacity $\kappa$ and photo-ionisation rate $\Gamma$ via the parameter $\xi$ such that $\kappa\propto\Gamma^{-\xi}$, and pose joint limits on \mfp and $\xi$. For the nominal value of $\xi=2/3$, we constrain $\lambda_{\text{mfp}} > 0.31 \ (0.18)$ pMpc at $2\sigma \ (3\sigma)$: a much tighter lower bound than obtained through traditional stacking methods. Our constraints get significantly stronger for lower values of $\xi$. New constraints on \mfp and $\xi$ are crucial to our understanding of the reionisation-era IGM.
\end{abstract}



\section{Introduction} \label{sec:intro}

The reionisation of hydrogen in the inter-galactic medium (IGM) is driven by ionising radiation emitted by the first stars. The unfolding of reionisation is therefore tightly linked to the evolving ionising emissivity of the first galaxies, but is also regulated by the average distance that ionising photons can travel through the IGM - the ionising mean free path $\lambda_{\text{mfp}}$. 

A fast increase in \mfp is an expected marker of the end of the reionisation process, indicating that the ionising background becomes percolated \citep{Gnedin00,Gnedin06, Daloisio18,Kulkarni19,Keating20,Nasir20}. Recently, \citet{Becker21} reported the measurement of a very short $\lambda_{\text{mfp}} = 0.75_{-0.45}^{+0.65}$ pMpc at $z=6.0$ and its increase by a factor of $12$ by $z=5.1$. Such a fast evolution would constitute a smoking gun of reionisation's end, adding to mounting evidence that the process completes at $z\lesssim6.0$ (e.g.~\citealt{Becker15,Becker18,Bosman18,Bosman21-tau,Boera19,Jung20,Kashino20,Morales21}, etc). A mean free path of $<1$ pMpc at $z=6$ requires the ionising emissivity of early galaxies to be significantly higher than observed in $z\leq3$ galaxies \citep{Cain21,Davies21}. Constraints on the mean free path at $z\gtrsim6$, at the end stages of reionisation, are thus a crucial element for models of the high-$z$ IGM. 

The ionising mean free path has been measured with methods falling broadly in three categories. First, the mean free path and the photo-ionisation rate ($\Gamma$) of the ultra-violet background (UVB) can be inferred in a model-dependent manner from measurements of the mean Lyman-$\alpha$ optical depth, provided the global emissivity of sources is known \citep{Miralda90,Meiksin93,Haardt96,Faucher08}. At $z>5.5$, the origin of UVB fluctuations is currently still debated and the global emissivity is unknown, therefore constraints are likely to be more model-dependent than at $z<5$ \citep{Davies16, Daloisio18,Nasir20}. A second category of methods consists in measuring the decline of the average transmitted flux of quasars at wavelengths shorter than the Lyman limit ($\lambda=911.76$\AA). The mean free path is defined as the distance over which the ionising flux has declined by a fraction $1/e$, corresponding to an opacity $\tau=1$. The flux decrease beyond the Lyman limit is therefore a direct measurement of \mfp \citep{Prochaska09,Fumagalli13,Omeara13,Worseck14,Becker21}. 

In this paper, we introduce a new method which most closely resembles the third approach: constraining the propagation distance of ionising radiation from individual sources (the individual free paths) via the distributions of absorbers which limit the propagation of radiation \citep{Songaila10,Rudie13,Romano19}. In practice, the propagation of ionising photons is often halted by encounters with discrete Lyman-limit systems (LLS) with $\log N_{\text{HI}} \gtrsim 17.2$ cm$^{-2}$. The distance from a quasar to the nearest LLS therefore poses a stringent lower limit on the individual free path around the object. Whereas \citet{Songaila10} and \citet{Romano19} identify a quasar's nearest LLS via absorption at the Lyman limit, we instead use the fact that an LLS will saturate all $6$ of the lowest-energy Lyman-series transitions (Ly-$\alpha$ to Ly-$\zeta$). Our method therefore has the advantage of requiring no transmission at the Lyman limit in order to pose constraints. Indeed, the overlapping Lyman-series transmission at the Lyman limit is predicted to be exceedingly weak \citep{Becker21}.

We explain our method in more detail in Section~\ref{sec:ifp}. We conduct a demonstration of the measurement on a sample of $26$ quasars at $5.5\lesssim z\lesssim 6.5$ and we present the individual free path constraints in Section~\ref{sec:obs}. The resulting limits on individual free paths place the potential LLS in regions where the photo-ionisation rate is dominated by the quasars rather than the UVB. To account for this effect, we employ the general framework developed by \citet{Becker21} and give details in Section~\ref{sec:model}. The mean free path away from the quasar's influence then depends on $\Gamma$ via a free parameter $\xi$. In Section~\ref{sec:results}, we present joint constraints on \mfp and $\xi$ and compare with previous measurements. We conclude in Section~\ref{sec:ccl}.

Throughout the paper we assume a \citet{Planck20} cosmology with $H_0 = 67.74, \Omega_m = 0.3089$. Wavelengths always refer to the rest-frame unless explicitly stated. Comoving and proper distances are always labelled explicitly (e.g.~cMpc). $M_{1450}$ corresponds to the absolute magnitude at $\lambda=1450$\AA.

\section{Measuring individual free paths with the Lyman series}\label{sec:ifp}

\begin{figure}
    \hskip-1.5em
    \includegraphics[width=1.1\columnwidth]{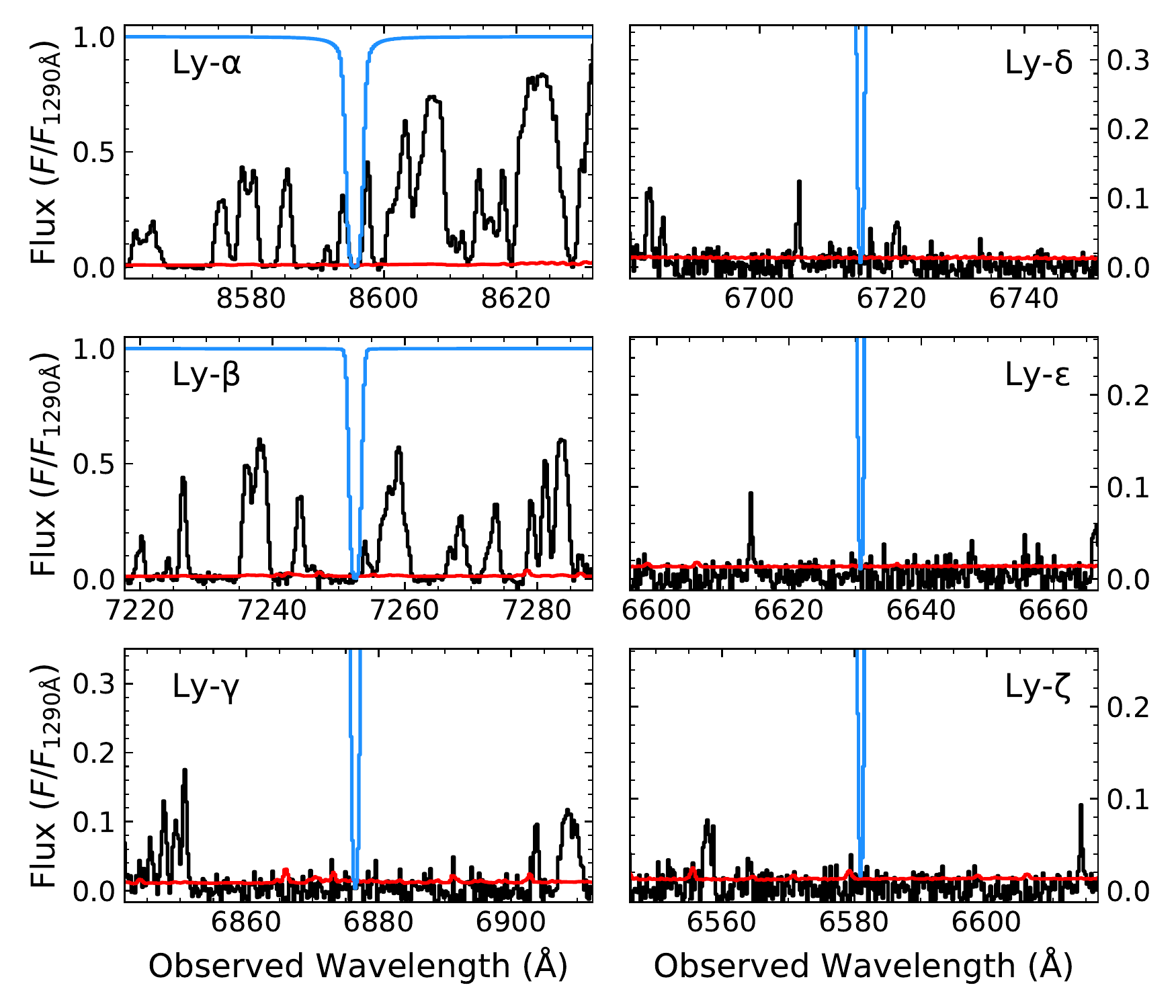}
    \caption{Example of the highest-$z$ possible LLS location in quasar J1319+0950. Note how saturated Ly-$\alpha$ absorbers do occur closer to the quasar (top left panel), but none of them are also opaque in Ly-$\beta$ and Ly-$\gamma$ (middle left and bottom left panels). The simulated absorbers have been forward-modelled to the resolution and pixel size of X-Shooter, assuming $b=15$ km s$^{-1}$. Note that the scale of the $y$-axis changes between figures.}
    \label{fig:example}
\end{figure}

The optical depth of the IGM to ionising radiation at a distance $r$ from a source of radiation is given by $\tau_{912} = -{\rm{ln}}(F(r)/F_0)$ in the rest frame of the absorbing gas, where $F_0$ is the ionising flux emitted by the source and $F(r)$ is the transmitted flux at distance $r$. 
Additionally, $\tau_{912} = \int_{r'=0}^r \kappa (r') dr'$, where $\kappa(r) = d\tau(r)/dr$. In a uniform medium, the definition of \mfp as the distance such that $\tau=1$ therefore leads to $\lambda_{\rm{mfp}} = 1/\kappa$.

In practice, the propagation of ionising radiation is halted by stochastic absorbers which individually have $\tau\gtrsim1$, corresponding to a column density of neutral gas $\log N_{\rm{HI}} /{\text{cm}}^{-2} \gtrsim 17.2$. In addition to suppressing ionising radiation at $\lambda<912$\AA, such absorbers will necessarily produce strong absorption in the lowest-energy Lyman transitions, i.e.~Ly-$\alpha \  1215.67$\AA, Ly-$\beta \ 1025.72$\AA, Ly-$\gamma \  972.53$\AA, Ly-$\delta \ 949.74$\AA, Ly-$\epsilon \ 937.80$\AA \ and Ly-$\zeta \ 930.74$\AA, to list the first $6$). For example, a neutral hydrogen absorber capable of saturating Ly-$\alpha$ but not Ly-$\beta$ absorption can have a column density no higher than ${\text{log}} N_{\rm{HI}} /{\text{cm}}^{-2} \simeq 15.5$. As seen in the lower-redshift Ly-$\alpha$ forest, even dozens of such absorbers do not result in significant absorption at the Lyman limit (e.g.~\citealt{Songaila10}, but see discussion of the contribution of absorbers with $N_{\rm{HI}} /{\text{cm}}^{-2} \gtrsim16.5$ in Sec.~\ref{sec:caveats}). For a LLS, absorption in the first $6$ Lyman transitions (at least) will occur in the non-linear regime of the curve of growth, resulting in flat central absorption troughs.


We use this property to find the nearest location from a quasar at which all $6$ transitions are strongly absorbed, directly corresponding to the \textit{nearest location at which a LLS could be located}. Straightforwardly, the individual free path of ionisation around the source can be no shorter than this limit. Overlap with foreground Ly-$\alpha$ absorption will lead, on average, to the limit being an under-estimate of the true individual free path; but it cannot be an over-estimate. 


The width of absorption troughs in the Lyman-series transitions depends on the Doppler parameter $b$. In the $z\sim6$ IGM, thermal broadening alone results in $b\geq10$ km s$^{-1}$ \citep{Gaikwad20}. However, inside the proximity zones where our limits on the individual free paths are located, photo-ionisation heating results in $b>20$ km s$^{-1}$ for $>80\%$ of hydrogen absorbers \citep{Bolton10,Bolton12}. We therefore opt for a nominal choice of $b=15$ km s$^{-1}$ (see Sec.~\ref{sec:caveats} for  further discussion of this choice). We generate Voigt profiles and forward-model them to the spectral resolution and pixel scales of the X-Shooter instrument, of $34$ km s$^{-1}$ and $10$ km s$^{-1}$ respectively \citep{XSHOOTER}. Spectroscopy of quasars with X-Shooter often achieves resolution superior to the nominal, closer to $28$ km s$^{-1}$ (see e.g.~\citealt{Bosman17}). We find that even at nominal resolution, the first $3$ Lyman series transitions are absorbed to less than $2\%$ transmission per pixel over the central $3$ pixels of the absorption troughs, and the next three transitions to less than $5\%$ transmission per pixel. We therefore select a criterion for candidate LLS locations of non-detections at the $3\sigma$ level over $3$ pixels, which determines the sensitivity level required of our observations (roughly SNR $\geq10$ per pixel). We neglect performing a reconstruction of the underlying continuum of the quasars, since obtaining accurate predictions (even within $30\%$) down to $930$\AA \ is highly impractical and limited by the availability of low-$z$ training samples \citep{Bosman21}.




\section{Observations}\label{sec:obs}

We use a sample of $26$ quasars at $5.5\lesssim z \lesssim 6.5$ observed with X-Shooter to a depth of SNR $\geq10$ per $10$ km s$^{-1}$ pixel (Table~\ref{tab:data}). All spectra were presented and used in the analysis of \citet{Bosman21-tau}, where more details can be found regarding the data reduction. The systemic redshifts of the quasars are obtained either from detections of the sub-mm emission lines of their host galaxies, or via the redshift of occurrence of the first IGM hydrogen absorber. The latter technique has an accuracy of $\Delta v = 180 \pm 180$ km s$^{-1}$ compared to the former \citep{Becker21}, corresponding to an uncertainty in the distance to the first candidate LLS location of $\Delta r = 0.2$ pMpc. We neglect this uncertainty.

\begin{table}[t]
    \centering
    \begin{tabular}{lllllc}
    Quasar & $z_{\text{qso}}$ & $M_{1450}$ & ${\text{max}} $ & ${\text{min}}\  \lambda_{\text{ifp}}$ & refs.\\
    name& & & $z_{\text{LLS}}$  & /pMpc & \\
    \hline\hline
J036+03 & $6.5405$ & $-27.33$ & $6.4847$ & $2.850$& (1,2)\\ 
J011+09 & $6.4695$ & $-25.95$ & $6.4227^{*}$ & $2.446$& (3,4)\\ 
J159$-$02 & $6.3860$ & $-26.80$ & $6.3584$ & $1.477$& (5,6)\\ 
J0100+2802 & $6.3269$ & $-29.14$ & $6.1955$ & $7.264$& (7,2)\\ 
J025$-$33 & $6.318$ & $-27.81$ & $6.2454^{*}$ & $4.001$& (8,9)\\ 
J1030+0524 & $6.309$ & $-26.99$ & $6.2425$ & $3.674$& (10,11)\\ 
J0330$-$4025 & $6.239$ & $-26.42$ & $6.2309$ & $0.452$& (12,13)\\ 
J308$-$21 & $6.2355$ & $-26.35$ & $6.2350$ & $0.024$& (5,2)\\ 
J2318$-$3029 & $6.1456$ & $-26.16$ & $6.1334$ & $0.706$& (14,2)\\ 
J1319+0950 & $6.1347$ & $-27.05$ & $6.0711$ & $3.730$& (15,2)\\ 
J1509$-$1749 & $6.1225$ & $-27.14$ & $6.1092$ & $0.778$& (16,17)\\ 
J2100$-$1715 & $6.0807$ & $-25.55$ & $6.0806$ & $0.001$& (18,2)\\ 
J1207+0630 & $6.0366$ & $-26.63$ & $6.0051$ & $1.902$& (19,17)\\ 
J1306+0356 & $6.033$ & $-26.81$ & $5.9190^{*}$ & $6.101$& (10,2)\\ 
J340$-$18 & $5.999$ & $-26.42$ & $5.9331$ & $0.007$& (20,9)\\ 
J0148+0600 & $5.998$ & $-27.39$ & $5.9437$ & $3.338$& (19,9)\\ 
J0818+1722 & $5.997$ & $-27.52$ & $5.9532$ & $2.706$& (21,9)\\ 
J0046$-$2837 & $6.021$ & $-25.42$ & $6.0201$ & $0.051$& (22,23)\\ 
J056$-$16 & $5.9676$ & $-26.72$ & $5.9551$ & $0.772$& (5,4)\\ 
J004+17 & $5.8166$ & $-26.01$ & $5.8164$ & $0.012$& (5,4)\\ 
J0836+0054 & $5.804$ & $-27.75$ & $5.4682^{*}$ & $22.888$& (10,6)\\ 
J0927+2001 & $5.7722$ & $-26.76$ & $5.7036^{*}$ & $4.586$& (21,23)\\ 
J215$-$16 & $5.7321$ & $-27.54$ & $5.7314$ & $0.043$& (24,24)\\ 
J1335$-$0328 & $5.693$ & $-27.76$ & $5.5938^{*}$ & $6.853$& (25,9)\\ 
J0108+0711 & $5.577$ & $-27.19$ & $5.4531^{*}$ & $8.975 $& (25,9)\\ 
J2207$-$0416 & $5.529$ & $-27.70$ & $5.4699$ & $4.326$& (26,9)\\ 
    \hline
    \hline
    \end{tabular}
    \caption{The quasars used in this work, with their physical and measured properties. Quasar redshifts with $5$ significant figures indicate the redshifts were obtained from sub-mm emission lines. Stars next to $z_{\text{LLS}}$ indicate that detection of significant Lyman continuum transmission refined the constraints ($7/26$ quasars). References correspond to (Discovery,Redshift): (1) \citealt{Venemans15}; (2) \citealt{Venemans20}; (3) \citealt{Mazzucchelli17}; (4)\citealt{Eilers20E}; (5) \citealt{Banados16}; (6)\citealt{Bosman21-tau}; (7) \citealt{Wu15}; (8) \citealt{Carnall15}; (9) \citealt{Becker19}; (10) \citealt{Fan01}; (11) \citealt{Jiang07}; (12) \citealt{Reed17}; (13) \citealt{Eilers20}; (14) \citealt{Farina19}; (15) \citealt{Mortlock09}; (16) \citealt{Willott07}; (17) \citealt{Decarli18}; (18) \citealt{Willott10}; (19) \citealt{Jiang15}; (20) \citealt{Banados15}; (21) \citealt{Fan06}; (22) \citealt{Venemans18}; (23) \citealt{Schindler20}; 
    (24) \citealt{Morganson12}; (25) \citealt{Yang17-qsos}; (26) \citealt{Wang16} }
    \label{tab:data}
\end{table}

Table~\ref{tab:data} lists the locations of the nearest possible LLS from each quasar. Figure~\ref{fig:example} shows an example of the first possible location of a LLS in quasar J1319+0950. Similar figures for all the quasars in our sample can be found in Appendix~\ref{sec:app1}.  For $7/26$ quasars, the constraints can be improved based on the occurrence of strong spikes of Lyman-continuum transmission at a lower redshift than our method alone. We conservatively define strong Lyman-continuum transmission as broad emission features detected at $>1\sigma$ over $6$ consecutive pixels, and at $>4\sigma$ overall. When Lyman continuum transmission is detected, we start the search for a LLS at $z\leq z_{\text{spike}}$ rather than $z\leq z_{\text{qso}}$, resulting in a lower max$(z_{\text{LLS}})$. The quasars for which Lyman continuum transmission provided an additional constraint are indicated by stars in Table~\ref{tab:data}. Figures showing the corresponding Lyman-continuum spikes for those $7$ quasars can be found in Appendix~\ref{sec:app2}.

\begin{figure*}
    \centering
    \includegraphics[width=1.0\textwidth]{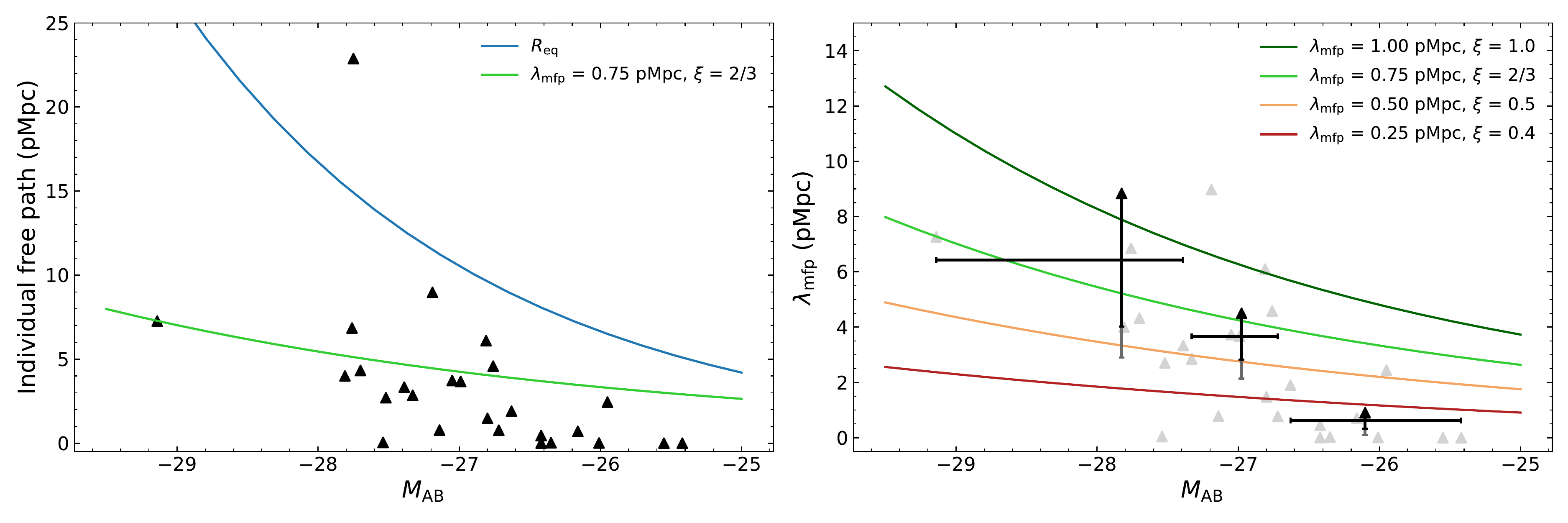}
    \caption{Left: Limits on individual free paths as a function of $M_{\rm{AB}}$ (black triangles) and expected proximity zone radius $R_{\text{eq}}$ (blue curve). Right: Limits on the mean free path and model curves, color-coded by degree of disagreement with the data (increasing green to red). The black bars represent the means of the individual free path limits, with the lower error bars encompassing $84\%$ (black) and $95\%$ (grey) of the bootstrap-resampled averages.}
    \label{fig:limits}
\end{figure*}

Figure~\ref{fig:limits} shows the resulting lower limits on individual free paths as a function of the quasar intrinsic magnitude $M_{1450}$. All of our lower limits on the individual free paths, except one, correspond to distances within the quasars' proximity zones with $r<R_{\text{eq}}$ (see Section~\ref{sec:model}). The sightline to quasar J0836+0054 is the sole exception. The lack of LLS absorption extends to a distance of $24.0$ pMpc in front of this object, $40\%$ longer than the quasar's expected influence given its $M_{1450}$. This suggests that J0836+0054, in addition to being highly luminous, happens to reside in a large-scale under-density (see also \citealt{Becker21}). Three quasars in our sample were targeted as potential `young' quasars with short Ly-$\alpha$ proximity zones \citep{Eilers20}. Two of them belong the lowest third of quasar magnitudes in our sample, among which one young quasar poses the weakest constraint (J2100$-$1715, tied with the non-young quasar J0046$-$2837) and one poses the tightest constraint on the individual free path (J011+09). We therefore find no qualitative difference between these quasars with the shortest proximity zones from the general population in terms of possible proximity to a LLS. We will return to this point in Section~\ref{sec:caveats}.

The trend of longer individual free paths with quasar brightness is a general theoretical expectation. Brighter quasars produce higher $\Gamma(r)$ in their environments, efficiently photo-ionising hydrogen and leading to longer individual free paths as well as fewer hydrogen absorbers in general.  Our model must therefore account for the quasars' ionising effects when computing the permitted \mfp in the IGM away from bright sources. We do not observe a trend in individual free path limits with redshift across our sample.




\section{Model}\label{sec:model}

In order to relate our limits on individual free paths around quasars to \mfp in the general IGM, we use the theoretically-motivated scaling of opacity with ionisation rate parametrised by a scaling parameter $\xi$:
\begin{equation}
    \kappa (\Gamma) = \kappa_{\text{bg}}  \left(\frac{\Gamma}{\Gamma_{\text{bg}}}\right)^{-\xi},
\end{equation}
where $\Gamma_{\text{bg}}$ is the ionisation rate in the IGM at $z=6$, which we set to $\Gamma_{\text{bg}} = 3 \times 10^{-13}$ s$^{-1}$ as measured by \citet{Becker21} (see Section~\ref{sec:caveats}). In a uniform and static medium, $\kappa_{\text{bg}}=1\lambda_{\text{mfp}}$. We adopt the general framework used by \citet{Becker21} to build a model which predicts the mean free path around a quasar of magnitude $M_{1450}$ as a function of the background \mfp and the scaling parameter $\xi$. 

The quasar sources a photo-ionisation rate which falls off quadratically with proper distance in an optically-thin medium:
\begin{equation}
    \frac{\Gamma_{\text{qso}}(r)}{\Gamma_{\text{bg}}} = \left(\frac{R_{\text{eq}}}{r}\right)^2,
\end{equation}
where $R_{\text{eq}}$ is the equilibrium distance  at which $\Gamma_{\text{qso}}(R_{\text{eq}}) = \Gamma_{\text{bg}}$ \citep{Calverley11}. To obtain $R_{\text{eq}}$, we use the scaling relation from \citet{Davies20-prox}:
\begin{eqnarray}
    R_{\text{eq}} = 11.3 \left( \frac{\Gamma_{\text{bg}}}{2.5 \times 10^{-13} {\text{s}}^{-1}}\right)^{0.5} \nonumber \\
    \times \left( \frac{\dot{N}_{\text{ion}}}{1.73\times 10^{57} {\text{s}}^{-1}} \right)^{-0.5} {\text{pMpc}},
\end{eqnarray}
where $\dot{N}_{\text{ion}}$ is the number of ionising photons emitted by the quasar per second. Finally, $\dot{N}_{\text{ion}}$ is obtained by extrapolating the spectral energy distribution of the quasar from $M_{1450}$ following a double power-law with $L(\nu) \propto \nu^{-\alpha_{\text{UV}}}$ at $912<\lambda<1450$\AA, and $L(\nu) \propto \nu^{-\alpha_{\text{ion}}}$ at $\lambda<912$\AA. We use $\alpha_{\text{UV}} = 0.6$ and $\alpha_{\text{ion}} = 1.5$ following \citet{Lusso15} (see Section~\ref{sec:caveats}).

To account for the absorption by residual neutral gas inside proximity zones, we solve equations (1) and (2) self-consistently by computing the radial profiles over small, optically-thin steps in radius $\Delta r$ (see \citealt{Davies14,Davies20-ghost,Becker21}):
\begin{equation}
    \tau(r+\Delta r) = \tau(r) + \frac{1}{\lambda_{\text{mfp}}} \int_{r'=r}^{r'=r+\Delta r}  \left(1 + \frac{\Gamma_{\text{qso}}(r')}{\Gamma_{\text{bg}}}\right)^{-\xi} dr';
\end{equation}
\begin{equation}
    \Gamma_{\text{qso}}(r+\Delta r) = e^{-\tau(r+\Delta r)}\Gamma_{\text{bg}} \left(\frac{R_{\text{eq}}}{r+\Delta r}\right)^2.
\end{equation}
We finally obtain \mfp via
\begin{equation}
    1 = \int_{r=0}^{r=\lambda_{\text{mfp}}} \kappa(r) dr.
\end{equation}
We refer the interested reader to \citet{Becker21} for more details of the framework. 

The resulting curves of \mfp as a function of $M_{1450}$ and $\xi$ are shown in Figure~\ref{fig:limits}. In general, models with high $\xi\geq2/3$ and \mfp$\geq0.75$ pMpc are consistent with $>50\%$ of individual free path limits at all magnitudes, which suggests that they are permitted by the observations. Conversely, models with a short \mfp and small $\xi$ increasingly fail to account for the large fraction ($>80\%$) of bright $M_{1450}<-26.5$ quasars which do not allow for short individual free paths. 


To quantify the level of tension between our models and the individual free path constraints, we first divide the sample into $3$ magnitude bins containing roughly the same number of objects ($9,9$ and $8$ objects from the faintest bin to the brightest) as shown in Figure~\ref{fig:limits}. The mean free path is equal to the mean of individual free paths, which can be {\textit{no lower than}} the mean of our lower limits on individual free paths. We therefore compute the mean of the individual free path limits in each magnitude bin. We calculate the uncertainty on these ``minimum \mfp limits'' by bootstrap-resampling the objects $40000$ times within each bin. The lower limits which encompass $84\%$ and $95\%$ of the re-sampled distribution correspond to the $1\sigma$ and $2\sigma$ lower bounds on $\lambda_{\text{mfp}}$, respectively. The probability of a given \mfp can then be obtained by integrating the distribution of re-sampled means from zero to $\lambda_{\text{mfp}}$, i.e.~over the range of real \mfp values which are compatible with the model's prediction. Formally, this procedure amounts to calculating the probability of the model given the observations. Note that we cannot rule in favour of particular values of the parameters, but only establish which set of $\{\lambda_{\rm{mfp}},\xi\}$ are in tension with the limits on individual free paths.



\section{Results}\label{sec:results}

\begin{figure}
    \hskip-2em
    \includegraphics[width=1.1\columnwidth]{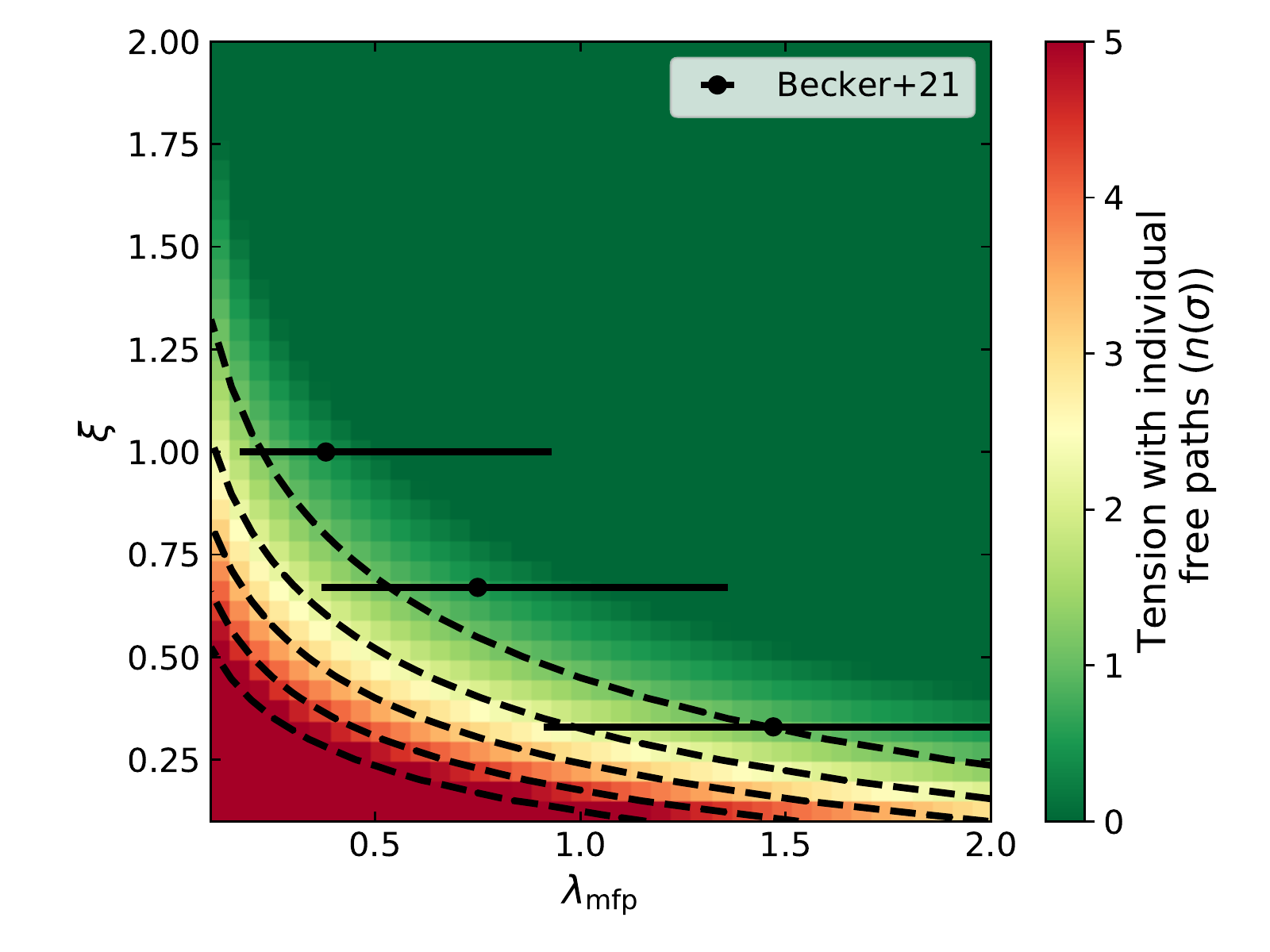}
    \caption{Posterior distribution of $\xi$ and $\lambda_{\text{mfp}}$ at $z\sim6$. The green-yellow-red color indicates the level of tension between the corresponding $\xi-\lambda_{\text{mfp}}$ model and the limits on individual free paths around quasars. The dashed curves delineate the bottom-left corner of parameter space, which is ruled out at $1,2,3,4,5 \sigma$ (curves in order of decreasing corner area). The \mfp measurements from \citet{Becker21} are shown for the authors' three different assumptions on $\xi$ ($0.33, 2/3, 1.0$).}
    \label{fig:result}
\end{figure}

Figure~\ref{fig:result} shows the level of tension between models and the individual free path constraints over parameter ranges $\lambda_{\text{mfp}} = [0.1,2.0]$ and $\xi = [0.1,2.0]$. The short \mfp value measured at $z=6$ by \citet{Becker21}, of \mfp$=0.75_{-0.45}^{+0.65}$, is permitted by our observations if $\xi=2/3$ as assumed nominally by the authors. The value of $\xi$ is tied to the physical properties of the LLS responsible for setting $\lambda_{\text{mfp}}$; $\xi=2/3$ is a theoretical expectation if their density profiles are isothermal \citep{Furlanetto05,McQuinn11}. For $\xi=2/3$, our observations pose constraints on the mean free path of $\lambda_{\text{mfp}}>0.53 \ (0.31,0.18)$ pMpc at $1\sigma\ (2\sigma,3\sigma)$. Our $2\sigma$ lower limits on \mfp are more constraining than those obtained from stacking transmission at the Lyman-limit ($\lambda_{\text{mfp}} > 0.1$ pMpc at $2\sigma$; \citealt{Becker21}).

Lower or higher values of the scaling parameter $\xi$ have been suggested in the literature. Based on hydro-dynamical simulations of dense self-shielded absorbers in the IGM, \citet{McQuinn11} and \citet{Daloisio20} obtain values of $\xi=0.75$ and $\xi=0.33$, respectively. Under the latter assumption for $\xi$, our observations pose tight lower limits of $\lambda_{\text{mfp}}>1.46 \ (1.00,0.68)$ pMpc at $1\sigma\ (2\sigma,3\sigma)$. These limits are again in agreement with the measurements of \citet{Becker21} under the same assumption, but our lower limits are more stringent (Fig.~\ref{fig:result}). For $\xi\geq0.75$, our method quickly becomes non-constraining because the expected \mfp around quasars becomes increasingly disconnected from the background $\lambda_{\text{mfp}}$.

Conversely, individual free paths can provide lower limits on $\xi$ given an assumption for $\lambda_{\text{mfp}}$. The reionisation models of \citet{Daloisio20} and \citet{Keating20-lyb} predict $2\lesssim \lambda_{\text{mfp}} \lesssim 4$ pMpc at $z=6$. Numerical simulations of the mean free path during reionisation have generally predicted a \mfp in the range of $1\lesssim \lambda_{\text{mfp}} \lesssim 6$ pMpc ($\lambda_{\text{mfp}}\sim6$ pMpc, \citealt{Alvarez12}; $\lambda_{\text{mfp}}\sim1.2$ pMpc, \citealt{Emberson13}; $4<\lambda_{\text{mfp}}<10$ pMpc, \citealt{Rahmati13}). The majority of these models are in tension with the nominal measurement of \citet{Becker21}; one way to ease the tension may be to invoke lower values of $\xi$ (Fig.~\ref{fig:result}). Our method becomes more constraining in the low-$\xi$ regime. Assuming a value of $\lambda_{\text{mfp}} = 2$ pMpc, our observations constrain a lower limit for $\xi>0.24\ (0.16,0.10)$ at $1\sigma\ (2\sigma,3\sigma)$. Using joint constraints on \mfp from stacking at the Lyman limit and individual free paths is therefore a promising way to study the physical state of the $z\sim6$ IGM.

\subsection{Caveats}\label{sec:caveats}

We discuss the impact of assumptions made in the analysis on our results, including the width of absorbers, uncertainties in model parameters, and the limitations of our formalism.

{\textit{Width of absorbers}:} Absorbers with Doppler broadening $b\lesssim12$ km s$^{-1}$ would not saturate the first $6$ Lyman transitions to the thresholds we have used in our measurements due to the relatively low resolution of X-Shooter. The temperature of the IGM at $z=5.8$ is $T\sim12,000$ K \citep{Gaikwad20}, corresponding to a floor of purely thermal broadening of $b \simeq10$ km s$^{-1}$ (but note that LLS are denser than the low-density IGM and thus potentially hotter). In a spectroscopically resolved analysis of hydrogen absorbers near quasars at $z\sim6$, \citet{Bolton10,Bolton12} instead found $b>20$ km s$^{-1}$ for over $80\%$ of Ly-$\alpha$ absorbers; the extra broadening is attributed to heating of the gas by the quasar to higher temperatures than the general IGM due to the ionisation of helium ($T\sim16000$K) and potentially additional kinematic broadening. Still, cold gas may conceivably be found inside quasar proximity zones, especially if the quasar phase started relatively recently and the helium-reionisation front has not yet reached the gas. To overcome this limitation, higher-resolution spectroscopy is necessary in order to resolve the widths of individual absorbers and check whether saturated absorption lines possess opaque troughs as expected. Unfortunately, samples of high-resolution $z>5.5$ quasar spectra do not yet exist in comparable numbers as the X-Shooter spectra used in this work. 

The assumed width of absorbers also relates to a caveat concerning the column densities of hydrogen absorbers which dominate the mean free path. Absorbers with $\log {N_\text{HI}}/$cm$^{-2} \leq 17.0$ have been argued to contribute significantly to limiting the propagation of ionising photons (e.g.~\citealt{Prochaska10,Haardt12,Rahmati18}) rather than \mfp being limited by the first encounter with a LLS. However, the same criteria of absorption in Ly-$\alpha$ through Ly-$\zeta$ employed in this work would be satisfied by an absorber with $\log {N_\text{HI}}/$cm$^{-2} \geq 16.8$ for $b = 20$ km s$^{-1}$, and $\log {N_\text{HI}}/$cm$^{-2} \geq 16.3$ for $b=25$ km s$^{-1}$. Our criteria therefore encompass the fact that the absorbers we located may be sub-LLS, making our limits conservative. In the future, we will employ numerical simulations to forward-model our individual free path procedure; the requirement for accurately predicting $b$ across a large range of densities necessitates novel radiation-tracing hydro-dynamical simulations which are beyond the scope of this work.


{\textit{Uncertainties in model parameters: }} We assume fixed values of $\alpha_{\text{UV}}=0.6$ and $\alpha_{\text{ion}}=1.5$ corresponding to the composite spectrum of \citet{Lusso15}. Both of these values are consistent with constraints from the wider literature (e.g.~$\alpha_{\text{UV}}$: \citealt{Berk01,Shull12}; $\alpha_{\text{ion}}$: \citealt{Telfer02}). The spread between different composite studies are $\Delta \alpha_{\text{UV}} \simeq 0.3$ and $\Delta \alpha_{\text{ion}} \simeq 0.5$. A systematic shift of both power-law indices to their $+1\sigma$ range would effectively result in a offset in our computed $M_{1450}$ of $\sim0.3$, insufficient to alter our results significantly. Differences in continuum properties among quasars will (effectively) lead to scatter in $M_{1450}$ which may alter the membership of individual sightlines to bins of emissivity; however, none of the quasars near the edges of the magnitude bins are exceptional compared to the neighbouring bin (see Fig.~\ref{fig:limits}). Confounding factors due to scatter in quasar emission properties can be alleviated by using larger samples in future work.

The uncertainty in $\Gamma_{\text{bg}}$ is potentially large. We use the value $\Gamma_{\text{bg}} = 3 \times 10^{-13}$ s$^{-1}$ which was most recently computed by \citet{Becker21} using the Sherwood simulation \citep{Bolton17}. This value of $\Gamma_{\text{bg}}$ is calibrated to match the mean Ly-$\alpha$ optical depth measured by \citet{Bosman18} under the assumption of a spatially homogeneous UVB, and has an uncertainty of $\sim 40\%$. However, the Sherwood simulation does not reproduce the observed spatial scatter of Ly-$\alpha$ opacity at $z=6$ \citep{Bosman21-tau}, suggesting that $\Gamma_{\text{bg}}$ may be biased. The late reionisation models  presented in \citet{Kulkarni19} and \citet{Keating20} successfully match the optical depth scatter, and predict $\Gamma_{\text{bg}} \simeq 2 \times 10^{-13}$ s$^{-1}$, $\Gamma_{\text{bg}} \simeq 3 \times 10^{-13}$ s$^{-1}$ respectively (without quantified uncertainties). A change in $\Gamma_{\text{bg}}$ would directly propagate to our measurements as $\lambda_{\text{mfp}} \propto \Gamma_{\text{bg}}^\xi$. Our analysis will therefore require revision in the event of significant updates to constraints on $\Gamma_{\text{bg}}$ at $z=6$.

{\textit{Limitations of the formalism: }} It is possible for the gas opacity around different quasars to follow different values of $\xi$. Dense gas inside proximity zones which was recently ionised by the quasar remains in a significantly non-relaxed state for $\gtrsim10^{4}$ years after quasar turn-on. Significant non-relaxation may lead to a lower effective value of $\xi$ compared to quasars which have been on continuously for $\gtrsim 10^7$ years (\citealt{Daloisio20}; see also \citealt{Becker21}). Indeed, two of the quasars in our sample have been argued to have particularly short current lifetimes $t<10^4$ years based on the short extent of their proximity zones \citep{Eilers21}. If quasar flickering on such timescales is common, the state of gas within proximity zones may not be well described by a single choice of $\xi$. We note, however, that (1) the two `young' quasars included in this work provided limits of individual free paths which were indistinguishable from the quasars which are not `young' in our sample; and (2) such an effect would only weaken our lower limits, making our parameters constraints overly conservative. Theoretical explorations of the effect of a flickering quasar phase on the surrounding gas are an active area of research (e.g.~\citealt{Davies20-prox,Chen20}); integrating non-equilibrium effects into a model of the IGM is beyond the scope of this work. We also neglect contributions to the ionisation rate from the galaxy populations which are expected to cluster around high-$z$ quasars. This contribution is expected to have a moderate impact \citep{Davies20-ghost}, but is non-trivial to model since the dark matter host halo masses of $z\sim6$ quasars are not currently known (see e.g.~\citealt{Habouzit19}).

In addition, we stress that the mean free path computed within the $\xi$ formalism always refers specifically  to the mean free path of ionising photons inside of ionised regions. Spatial fluctuations of $\Gamma_{\text{bg}}$ in the IGM, such as those induced by `islands' of significantly neutral gas, are not included in the model and cannot impact the resulting constraints on $\lambda_{\text{mfp}}$; in other words, \mfp is assumed to not arise from reionisation morphology. Comparisons of constraints obtained within the framework with models in which morphological effects drive \mfp is therefore non-trivial.



\section{Conclusions}\label{sec:ccl}

We have introduced a new method for constraining the mean free path of ionising photons at $z=6$ using lower limits on the individual free paths. We use the fact that Lyman-limit absorbers with a density sufficient to halt the propagation of ionising photons necessarily produce strong absorption in the $6$ lowest-energy Lyman transitions. In the absence of such features, the presence of a Lyman-limit system can be ruled out and a lower limit on the individual free path calculated.

We find that $\sim60\%$ of our sample of $26$ quasars require individual free paths $\lambda>2$ pMpc. In quasars with magnitudes brighter than $M_{1450}=-26.5$, the fraction rises to $83\%$ ($15/18$). Dividing our sample into three magnitude bins containing equal numbers of objects, we use the resulting lower limits on the mean free path around quasars to jointly constrain the mean free path in the background IGM and the scaling of the mean free path on opacity, $\xi$. We find constrains on \mfp which are in agreement with the measurements of a short \mfp by \citet{Becker21}, but our method poses more stringent lower limits on $\lambda_{\text{mfp}}$. For the traditionally-assumed value of $\xi=2/3$, we constrain $\lambda_{\text{mfp}}> 0.31 \ (0.18)$ pMpc at $2\sigma \ (3\sigma)$. Lower values of $\xi$ tighten the constraints: for $\xi=0.33$ as expected from ionisation non-equilibrium around quasars \citep{Daloisio20}, we constrain $\lambda_{\text{mfp}}> 1.00 \ (0.68)$ pMpc at $2\sigma \ (3\sigma)$. 

Our limits on \mfp are complementary with the approach of stacking transmission at the Lyman limit, as they require no transmission of Lyman continuum to be detected. Individual free path limits can be measured even in the case of overlap with strong Ly-$\alpha$ foreground absorption, as long as quasar possess proximity zones. Independent constraints on \mfp and $\xi$ will be crucial to understanding the IGM at the end stages of reionisation.

\begin{acknowledgments}
The author thanks George Becker for sharing his reductions of some of the quasar spectra used in this work and for constructive comments on the manuscript. The author is grateful for productive discussions and insightful feedback from Frederick Davies, Joe Hennawi and G\'{a}bor Worseck. 

SEIB acknowledges funding from the European Research Council (ERC) under the European Union's Horizon $2020$ research and innovation programme (grant agreement No.~$740246$ ``Cosmic Gas'').
\end{acknowledgments}

%






\appendix

\section{First possible locations of a Lyman-limit system}\label{sec:app1}

Figure~\ref{fig:all_transitions} the location of the first LLS absorptions in Ly-$\alpha$ through Ly-$\zeta$, based on non-detection criteria over three pixels in each transition as explained in the main text. For $7$ quasars where the presence of significant Lyman continuum transmission spikes helped to refine the constraints, we indicate the location of the transmission spike redshift with orange vertical bars.

\begin{figure*}
    \figurenum{4}
    \centering
    \includegraphics[width=0.49\columnwidth]{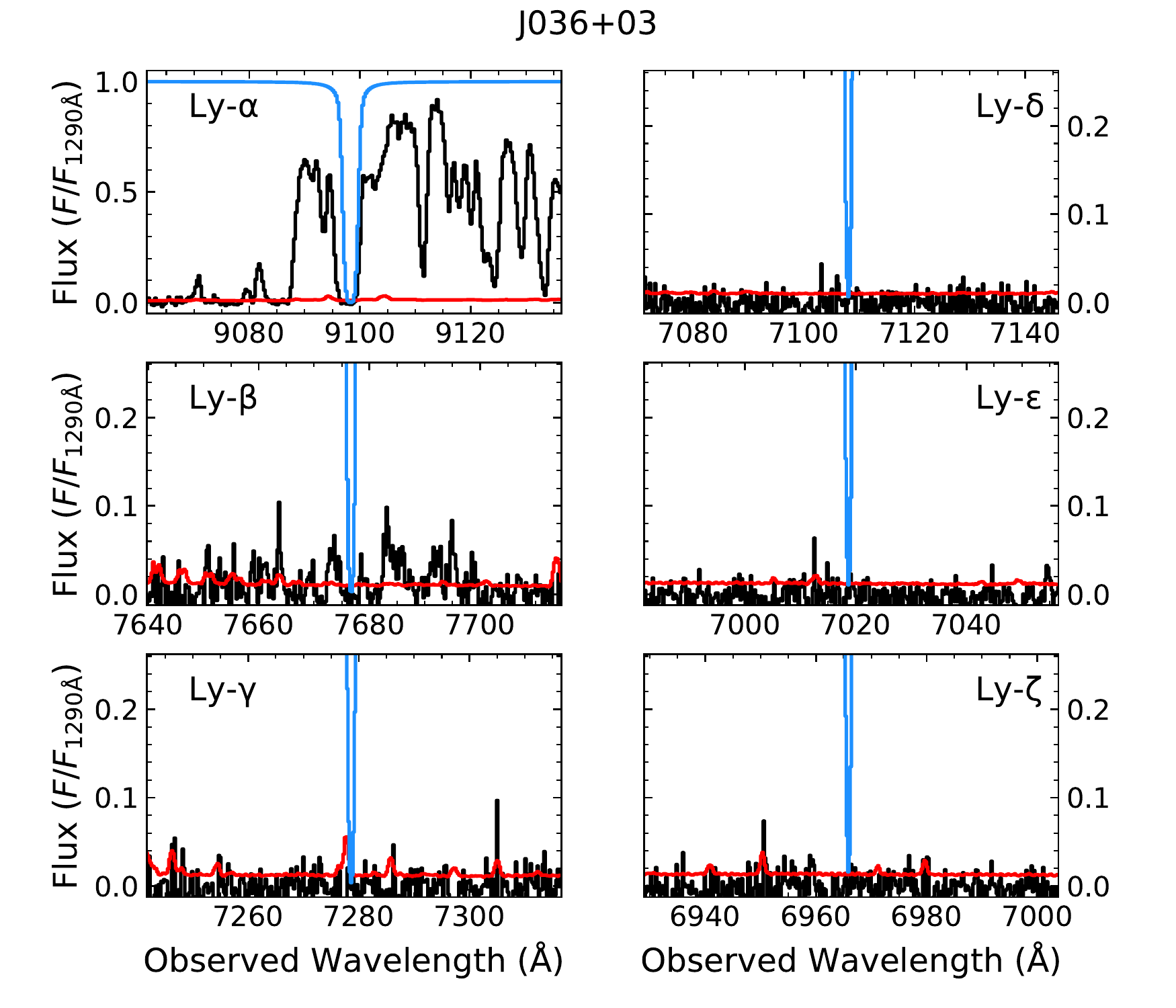}
    \includegraphics[width=0.49\columnwidth]{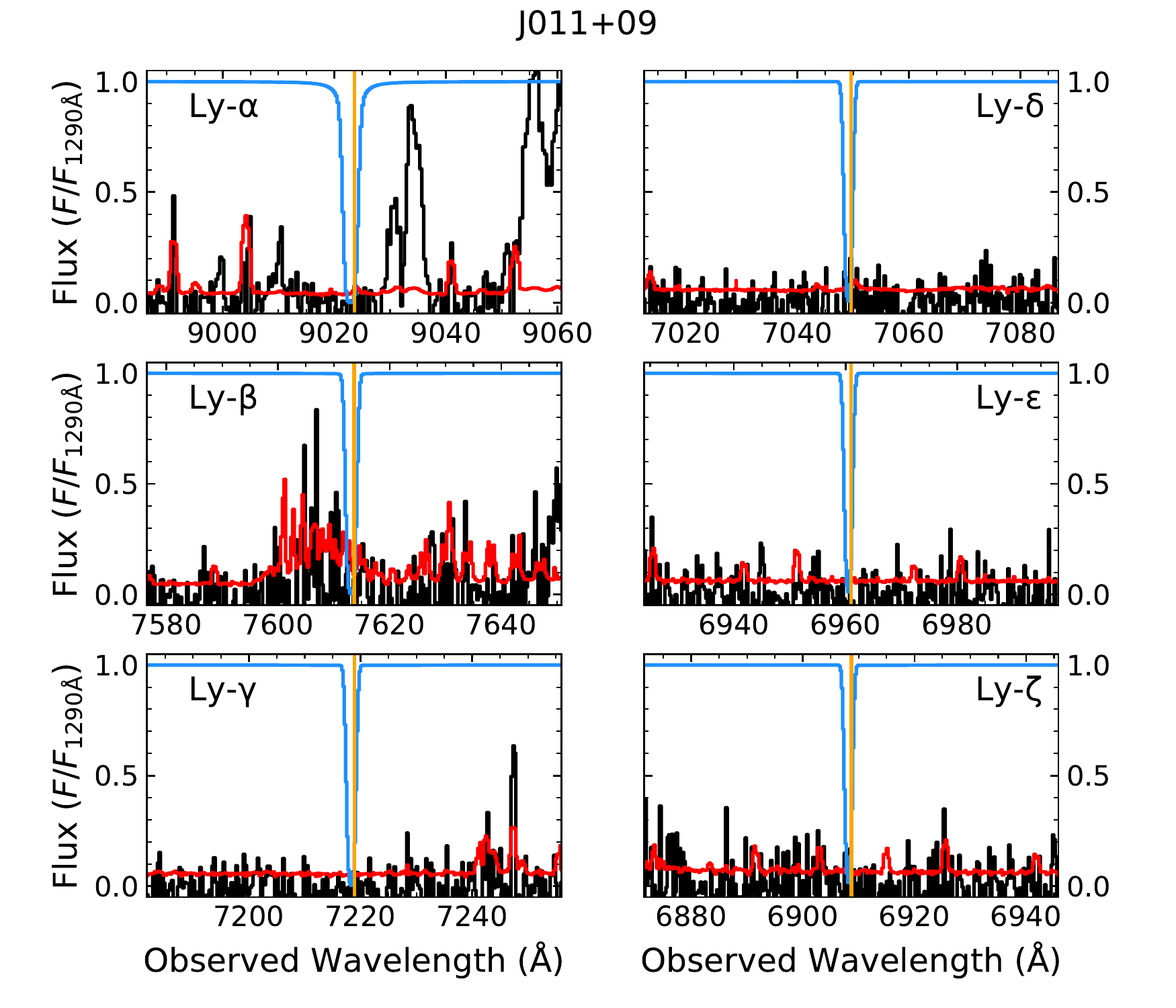}
    \includegraphics[width=0.49\columnwidth]{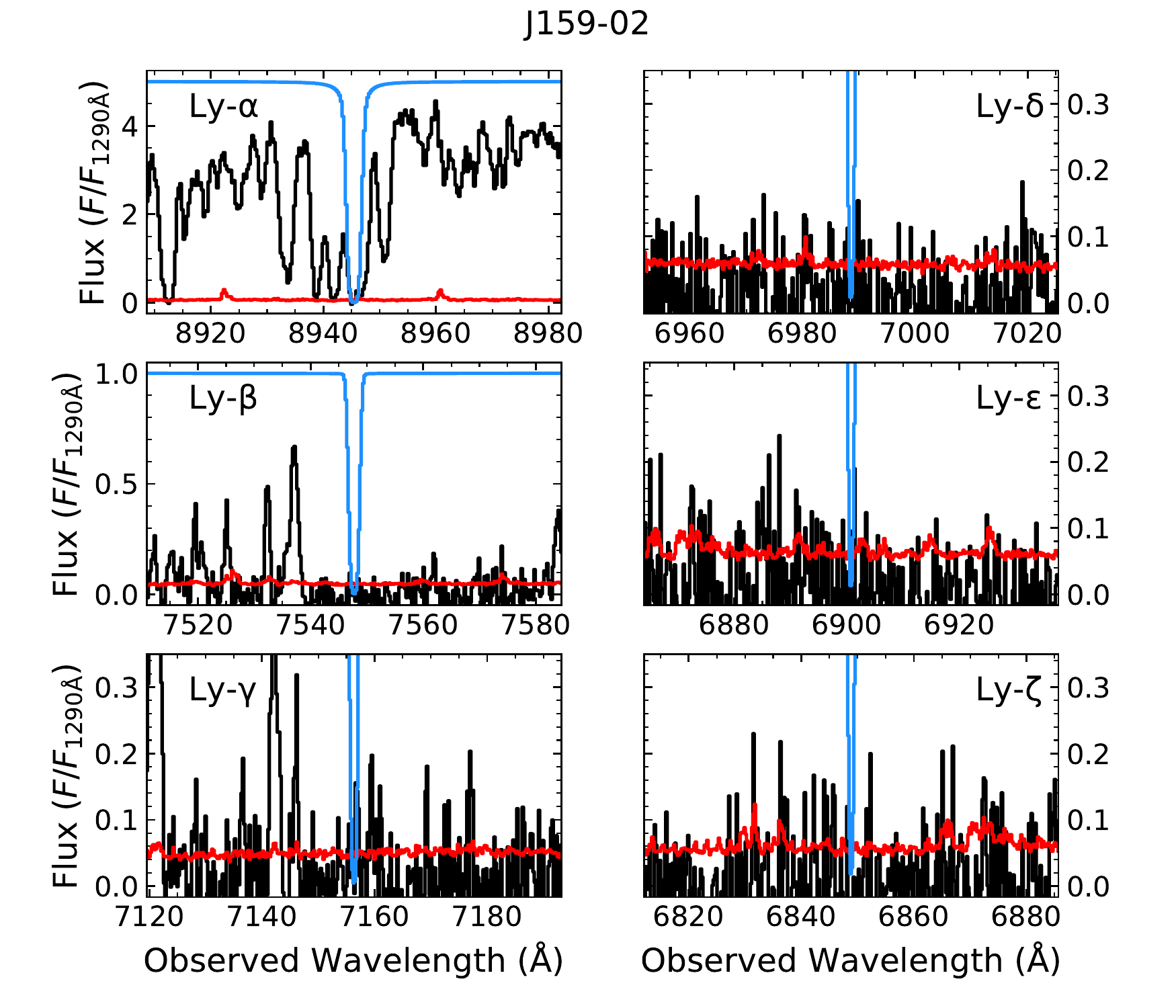}
    \includegraphics[width=0.49\columnwidth]{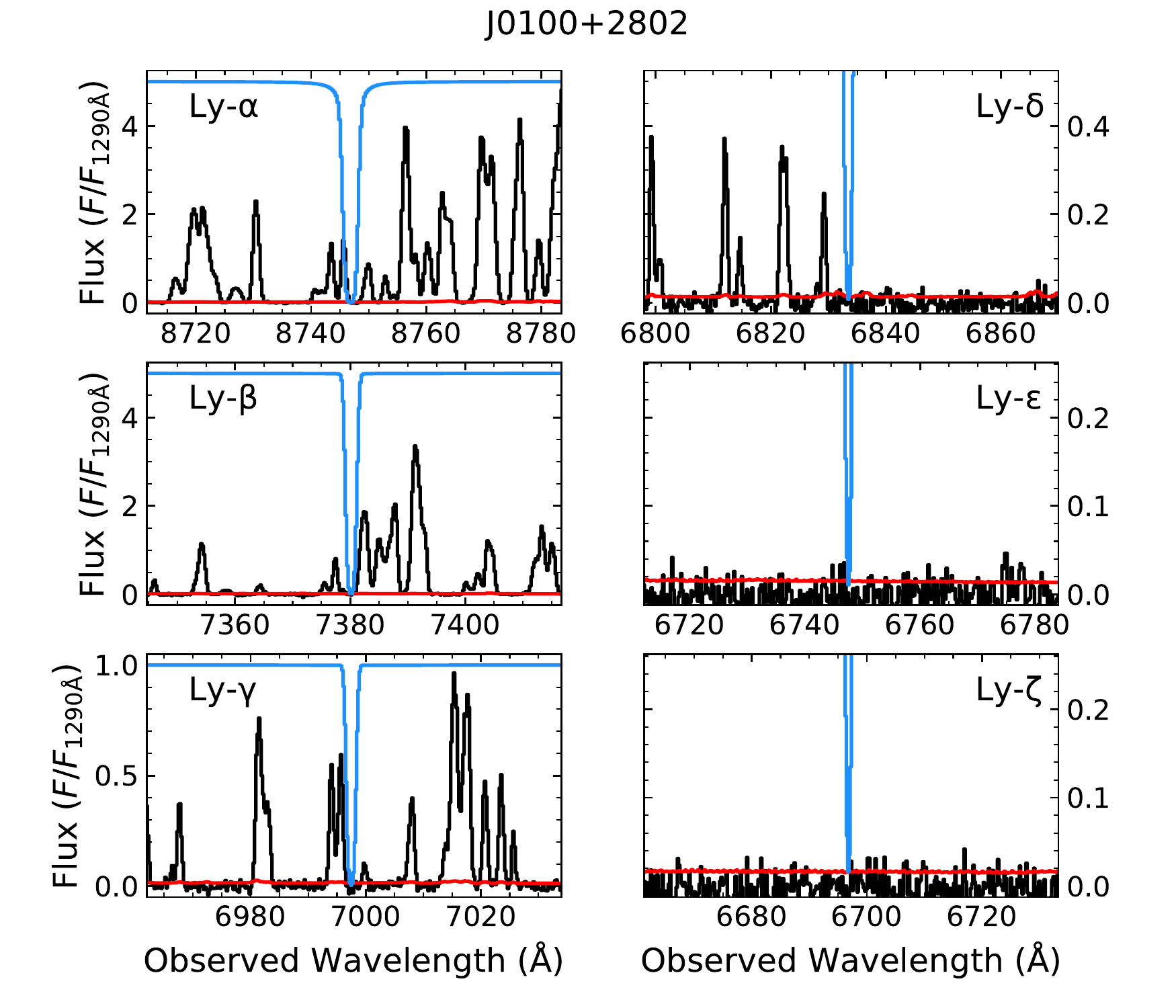}
    \includegraphics[width=0.49\columnwidth]{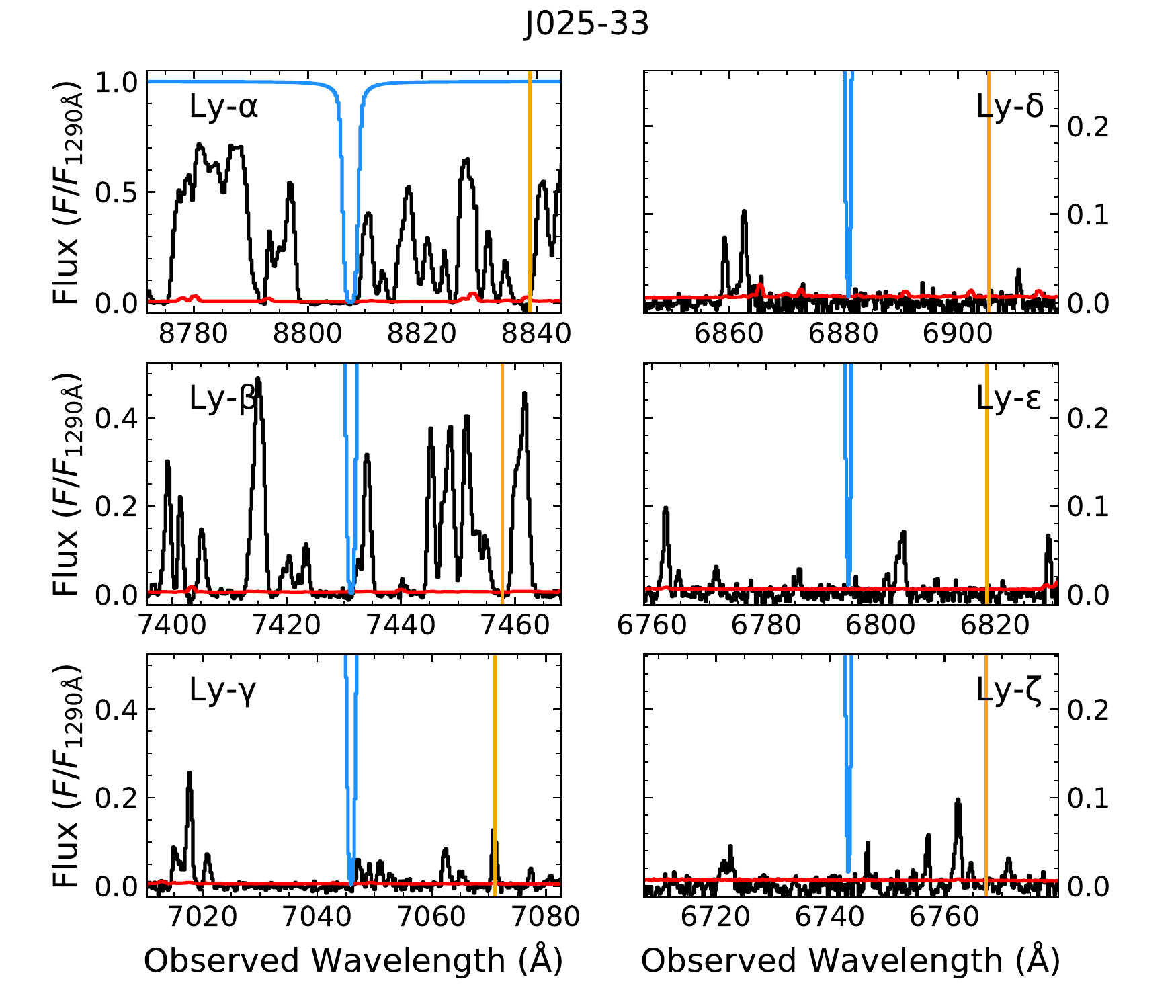}
    \includegraphics[width=0.49\columnwidth]{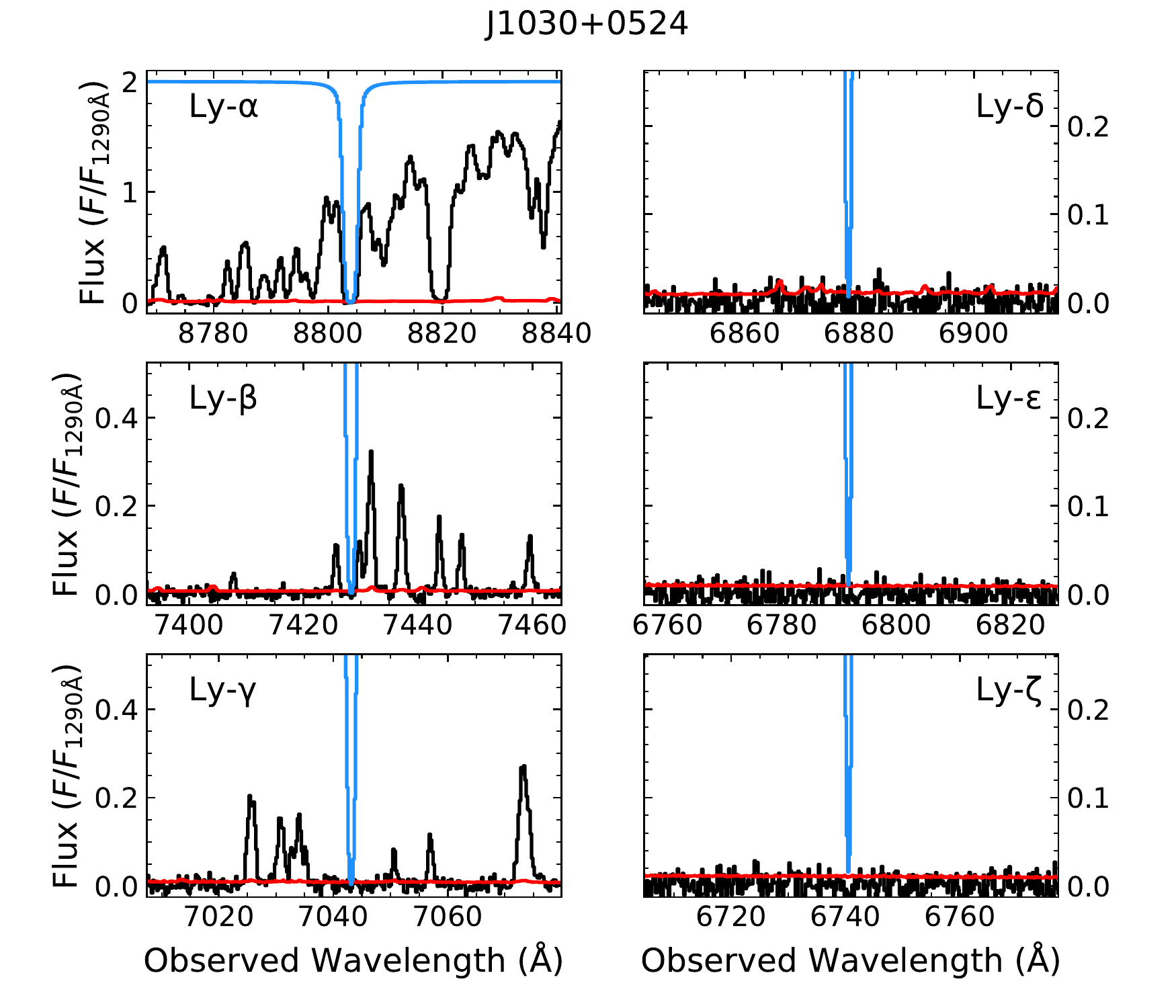}
    \caption{Same as Figure~\ref{fig:example} for all quasars in the sample, in order of decreasing redshift.}
    \label{fig:all_transitions}
\end{figure*}
\begin{figure*}
    \figurenum{4}
    \centering
    \includegraphics[width=0.49\columnwidth]{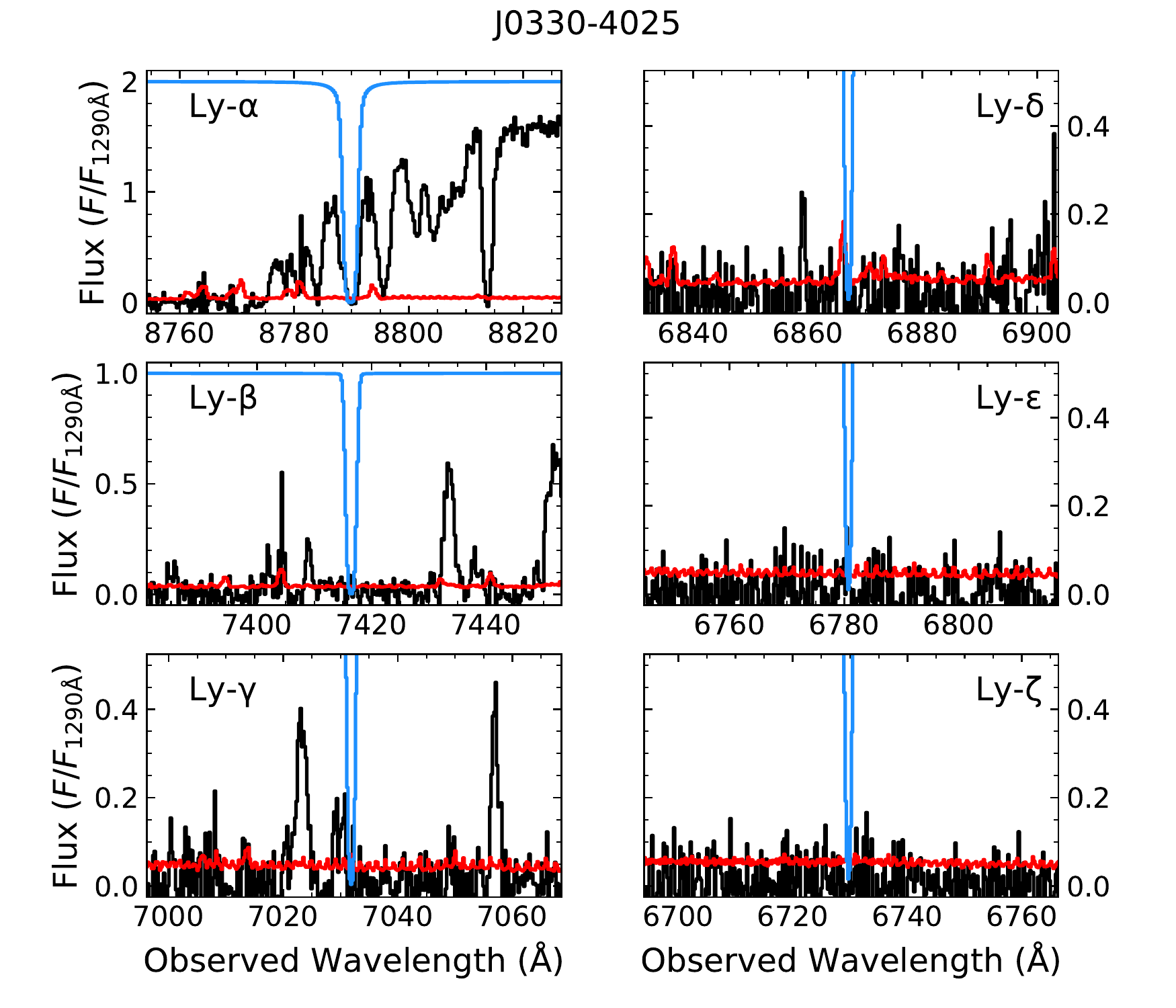}
    \includegraphics[width=0.49\columnwidth]{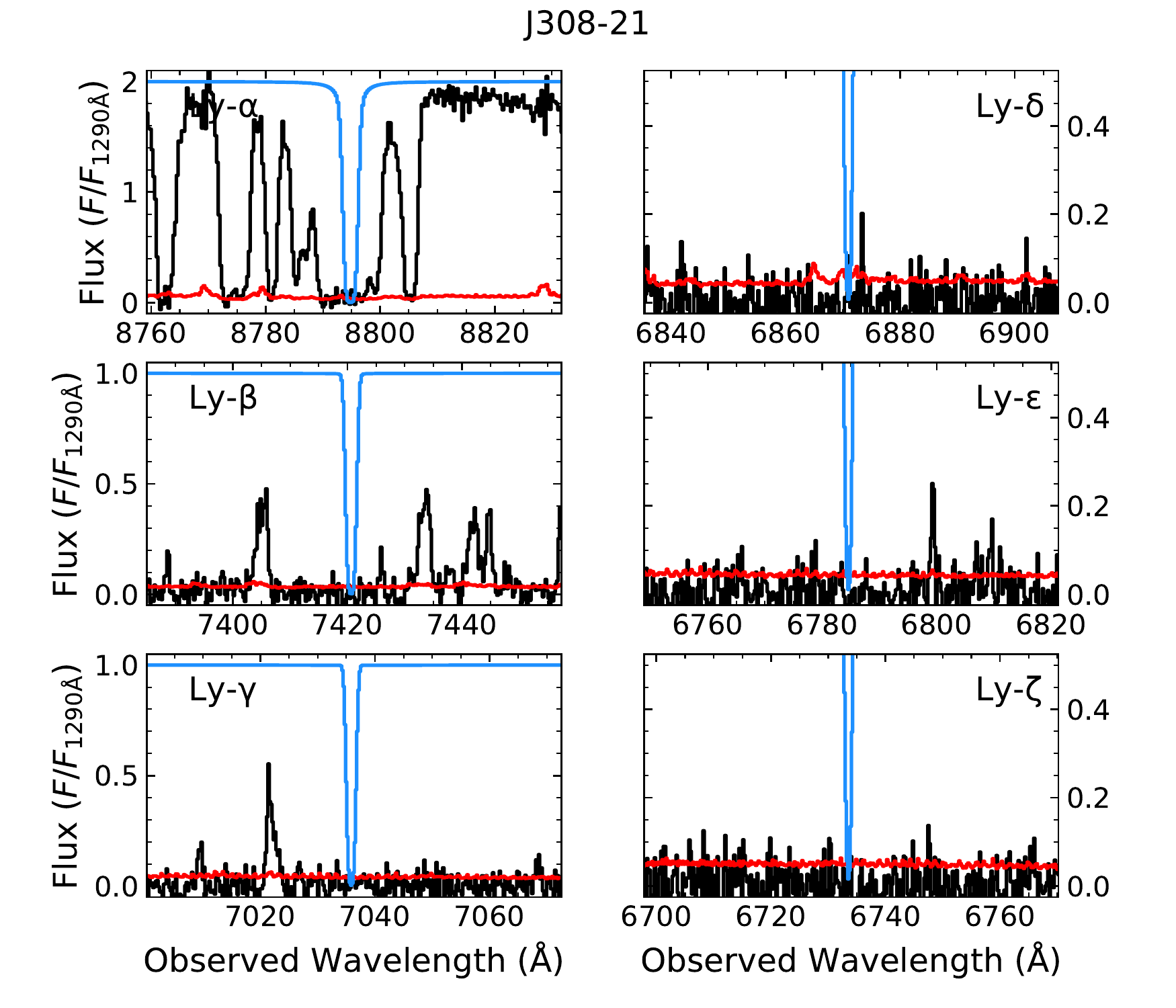}
    \includegraphics[width=0.49\columnwidth]{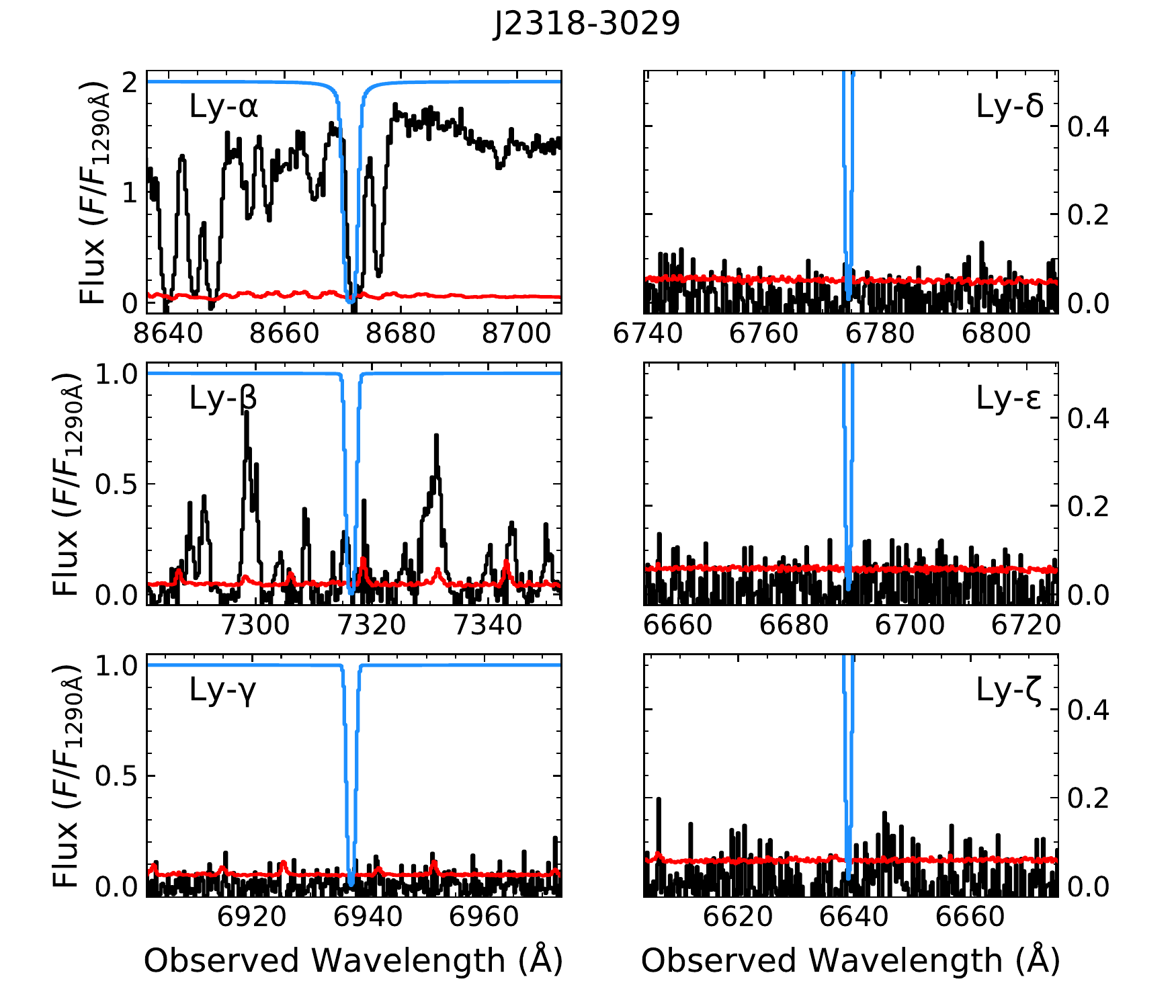}
    \includegraphics[width=0.49\columnwidth]{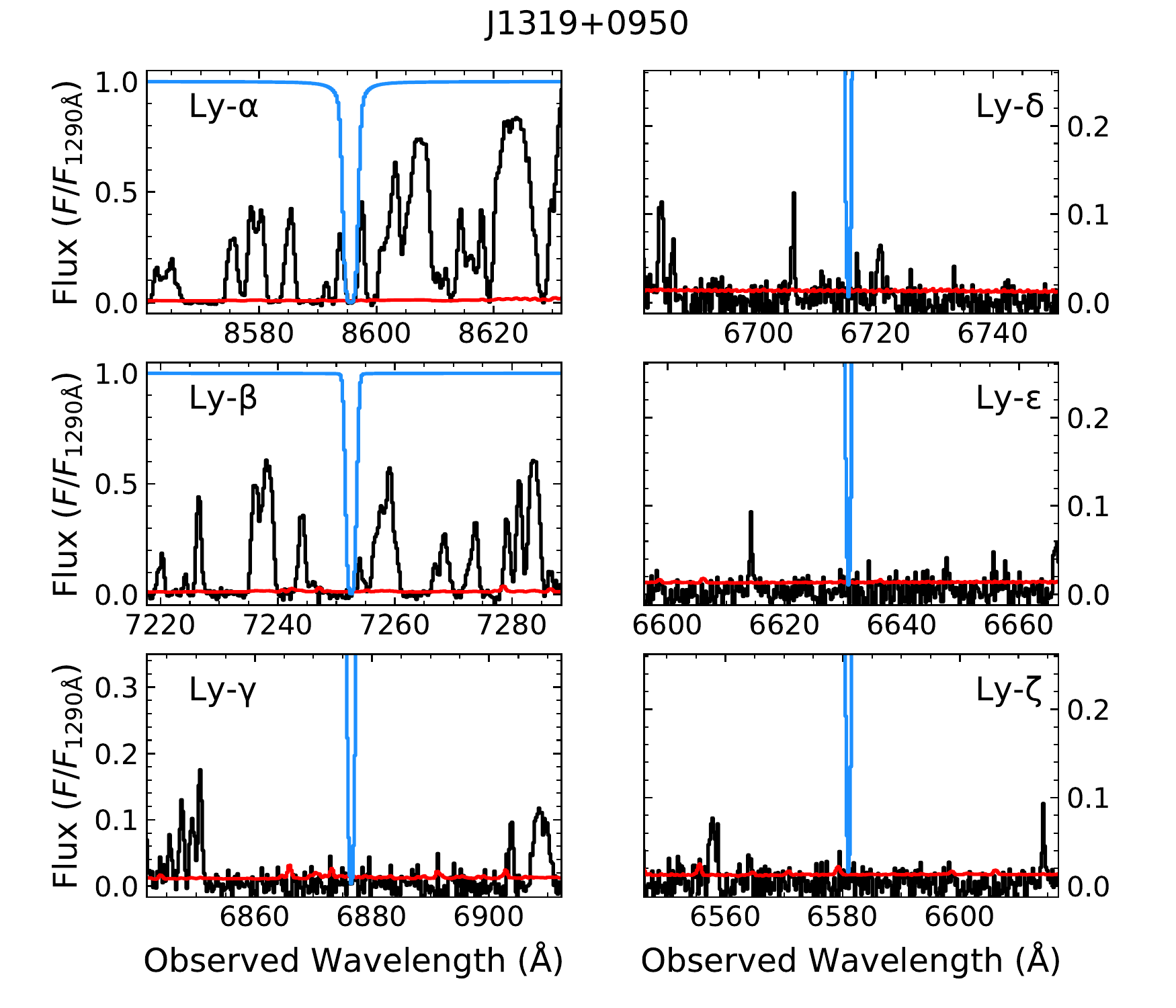}
    \includegraphics[width=0.49\columnwidth]{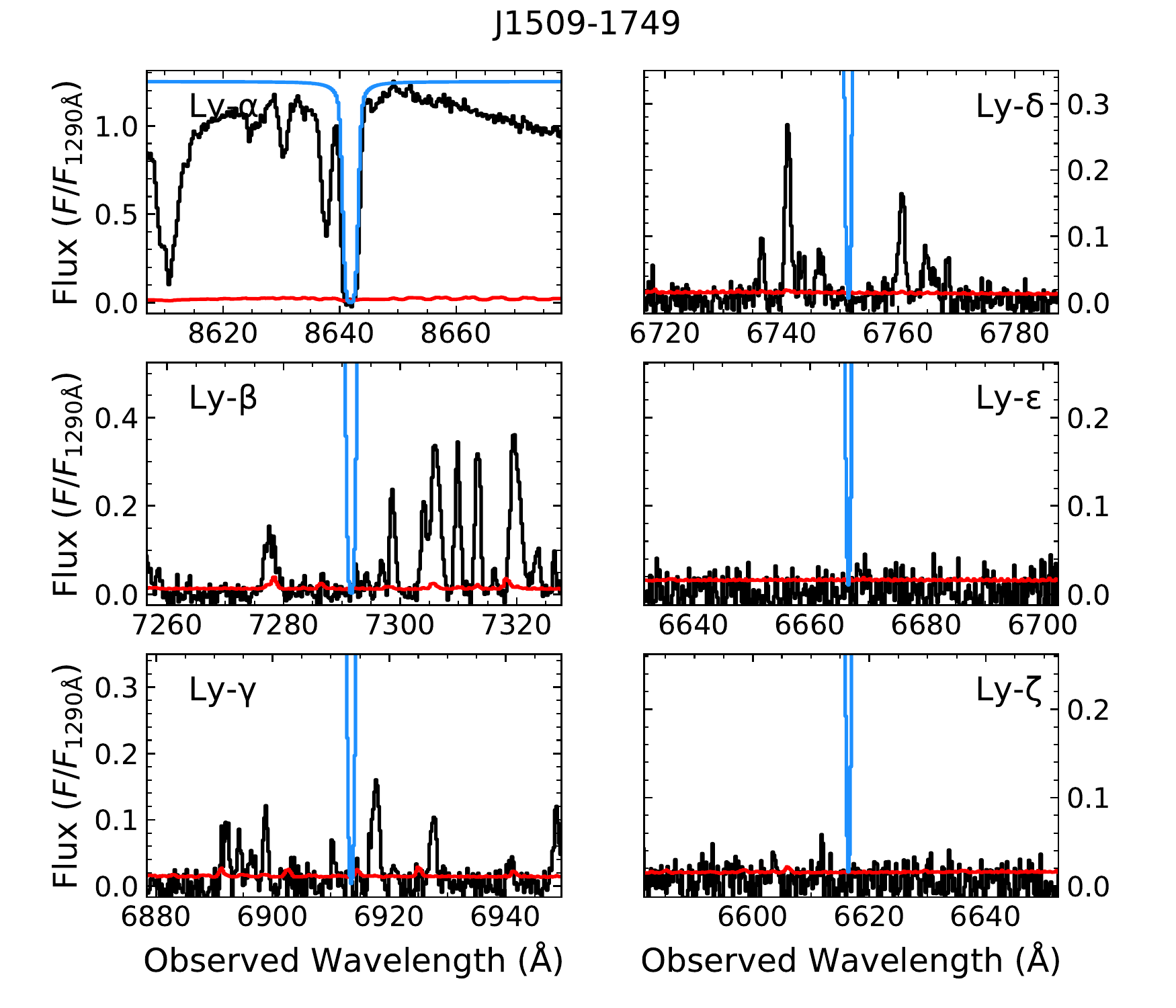}
    \includegraphics[width=0.49\columnwidth]{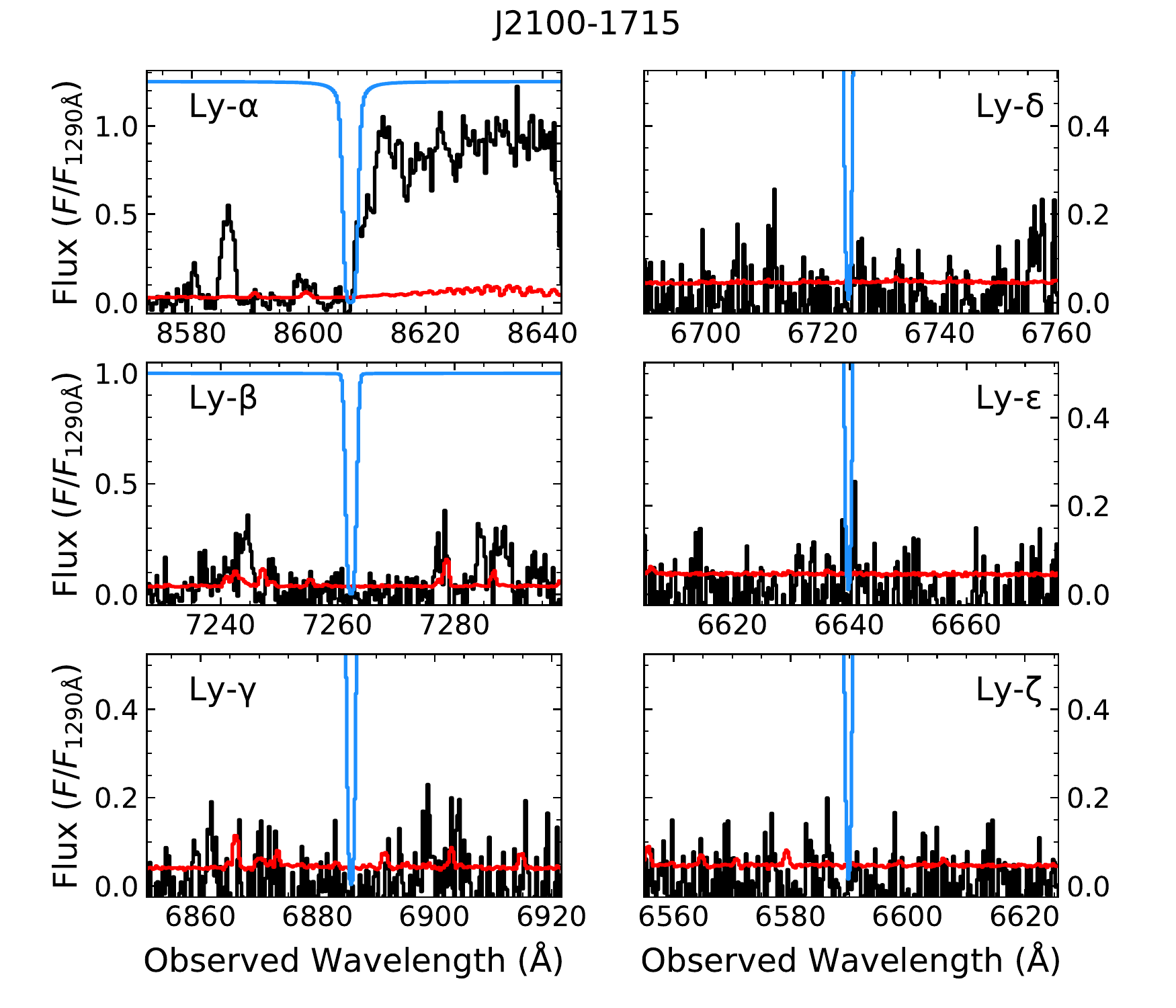}
    \caption{{\textit{continued}}}
\end{figure*}
\begin{figure*}
    \figurenum{4}
    \centering
    \includegraphics[width=0.49\columnwidth]{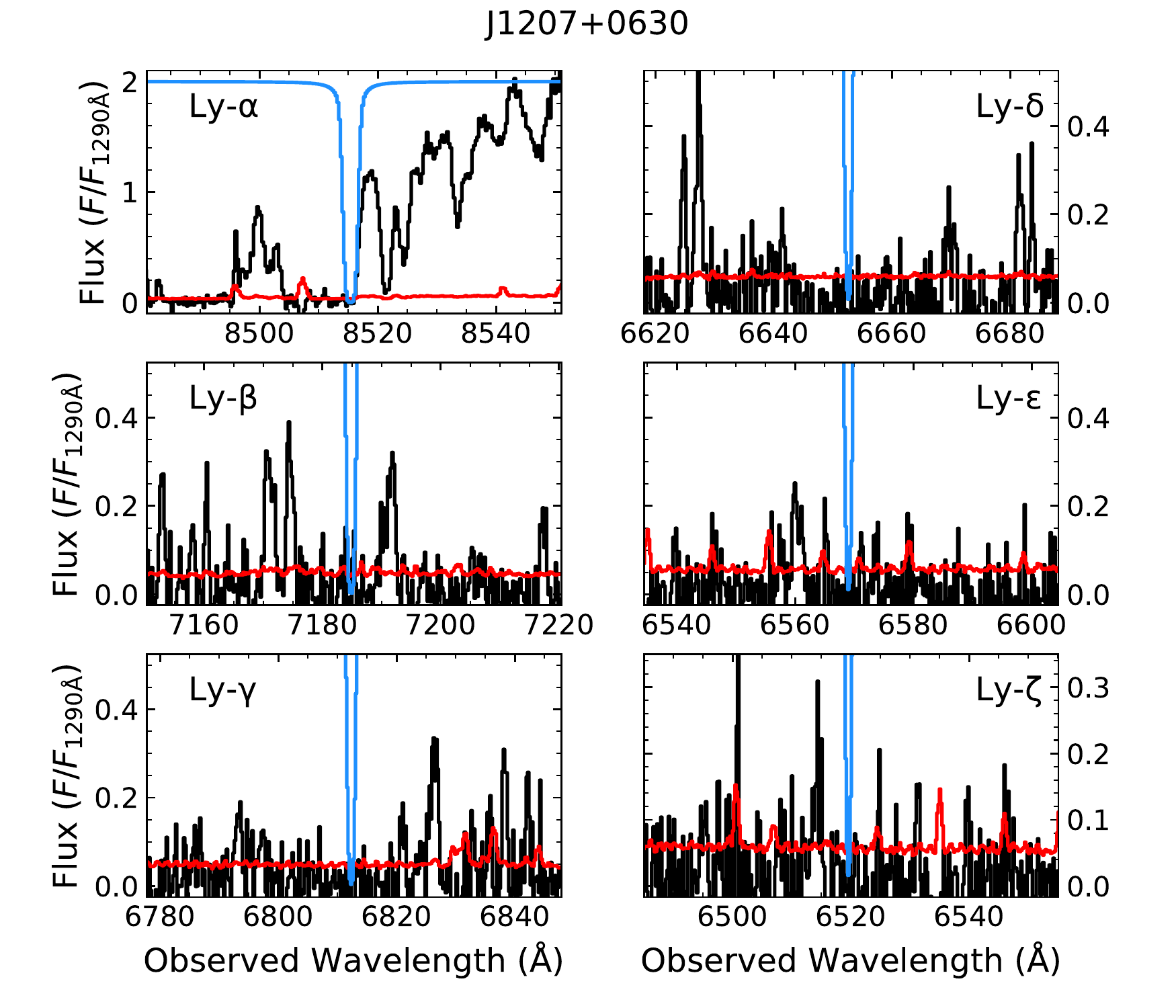}
    \includegraphics[width=0.49\columnwidth]{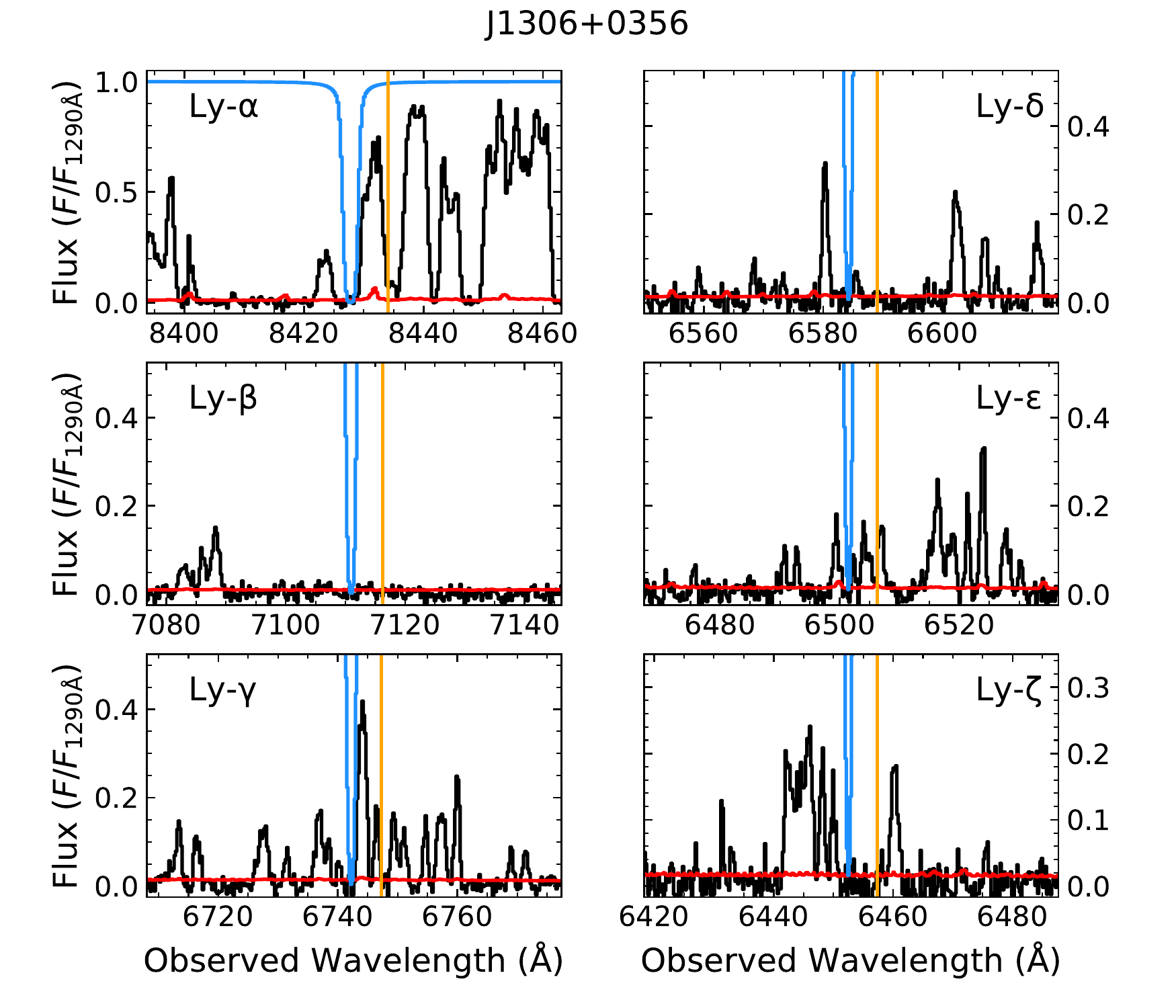}
    \includegraphics[width=0.49\columnwidth]{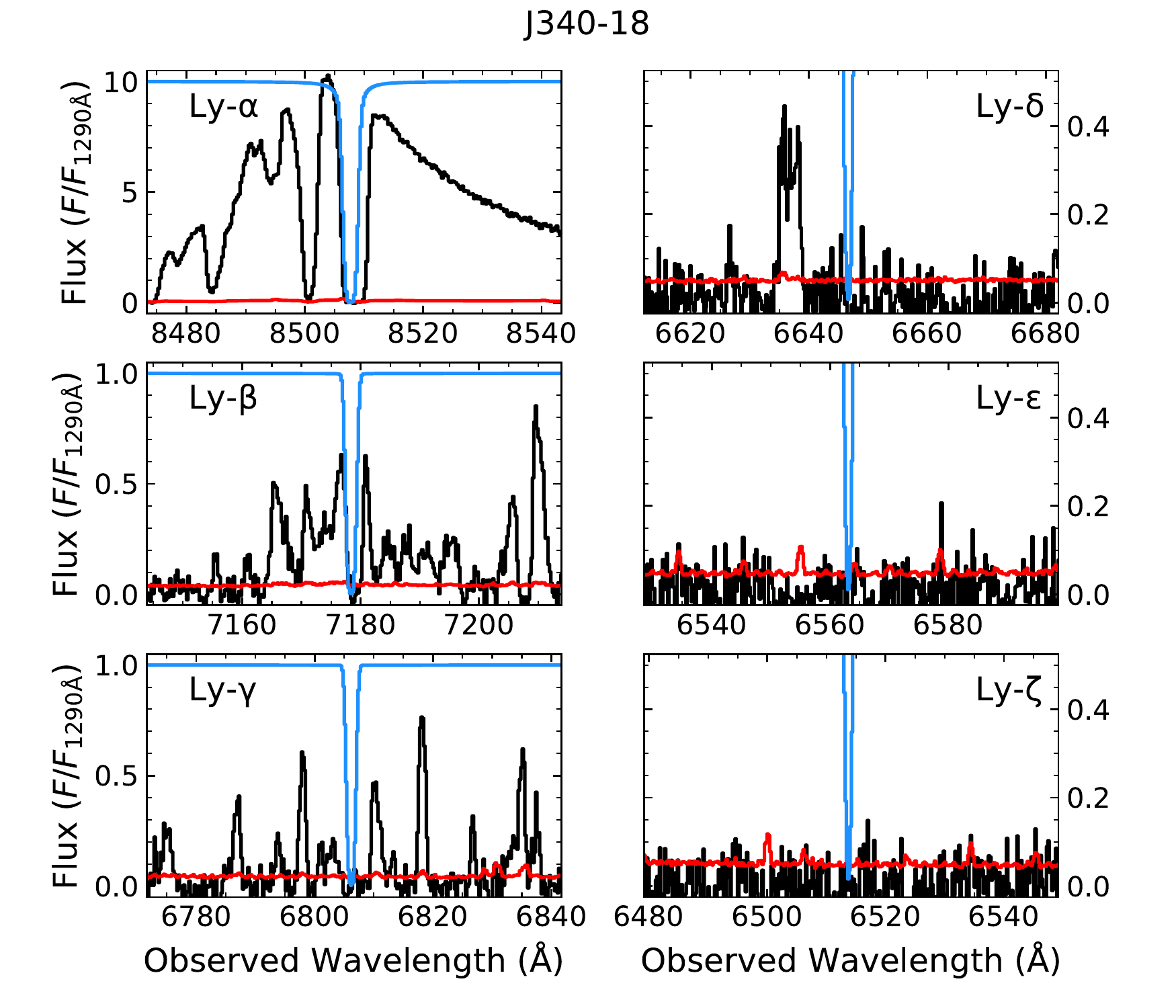}
    \includegraphics[width=0.49\columnwidth]{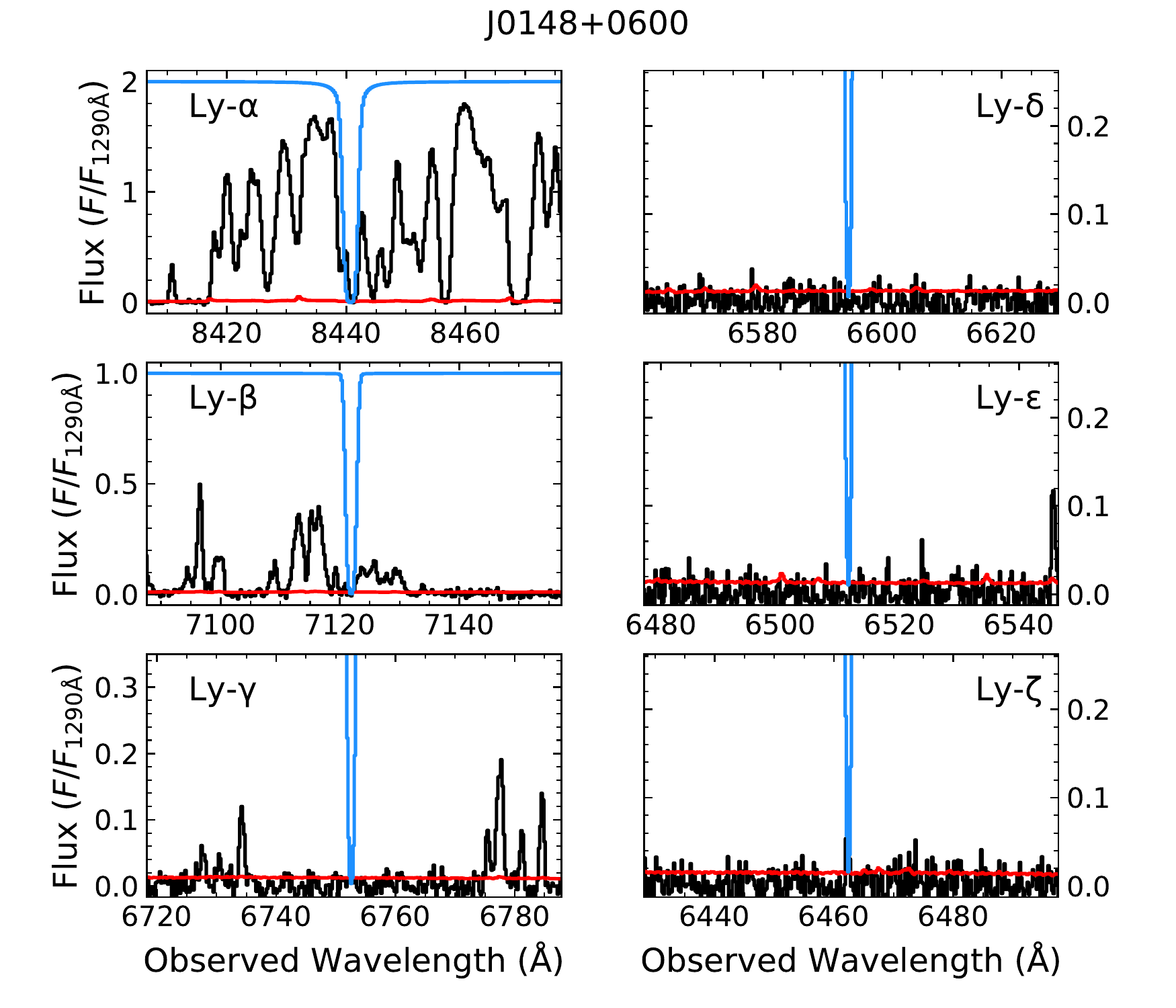}
    \includegraphics[width=0.49\columnwidth]{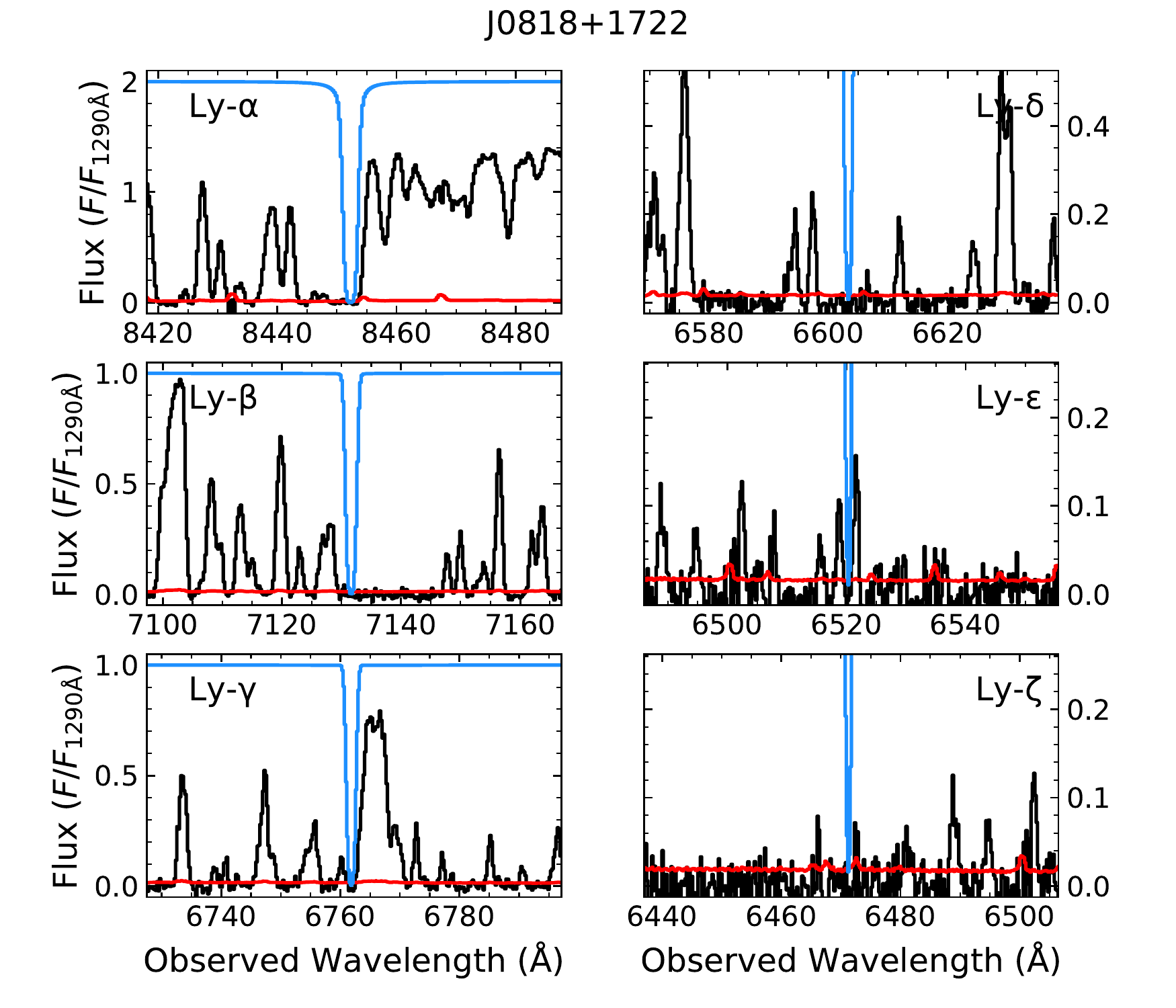}
    \includegraphics[width=0.49\columnwidth]{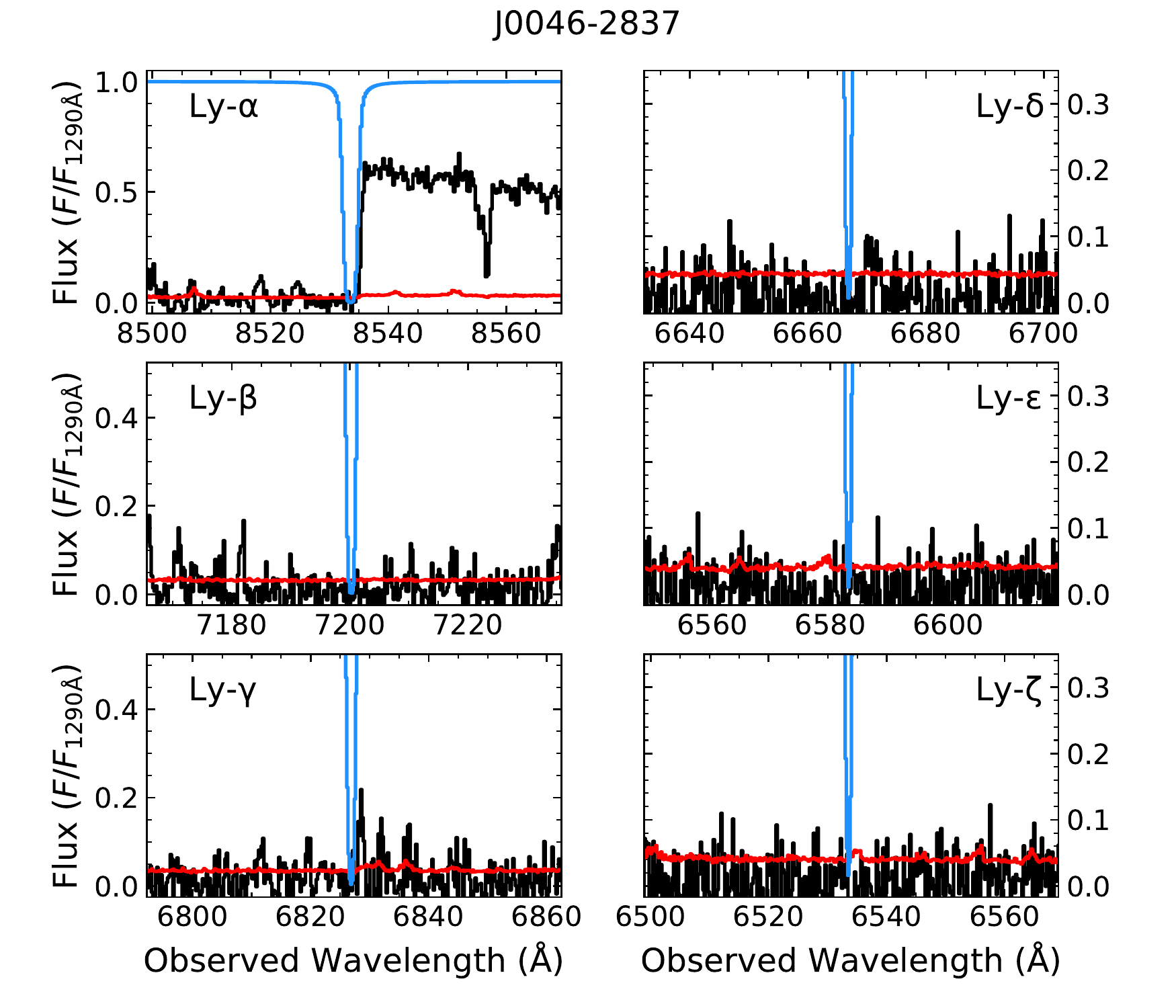}
    \caption{{\textit{continued}}}
\end{figure*}
\begin{figure*}
    \figurenum{4}
    \centering
    \includegraphics[width=0.49\columnwidth]{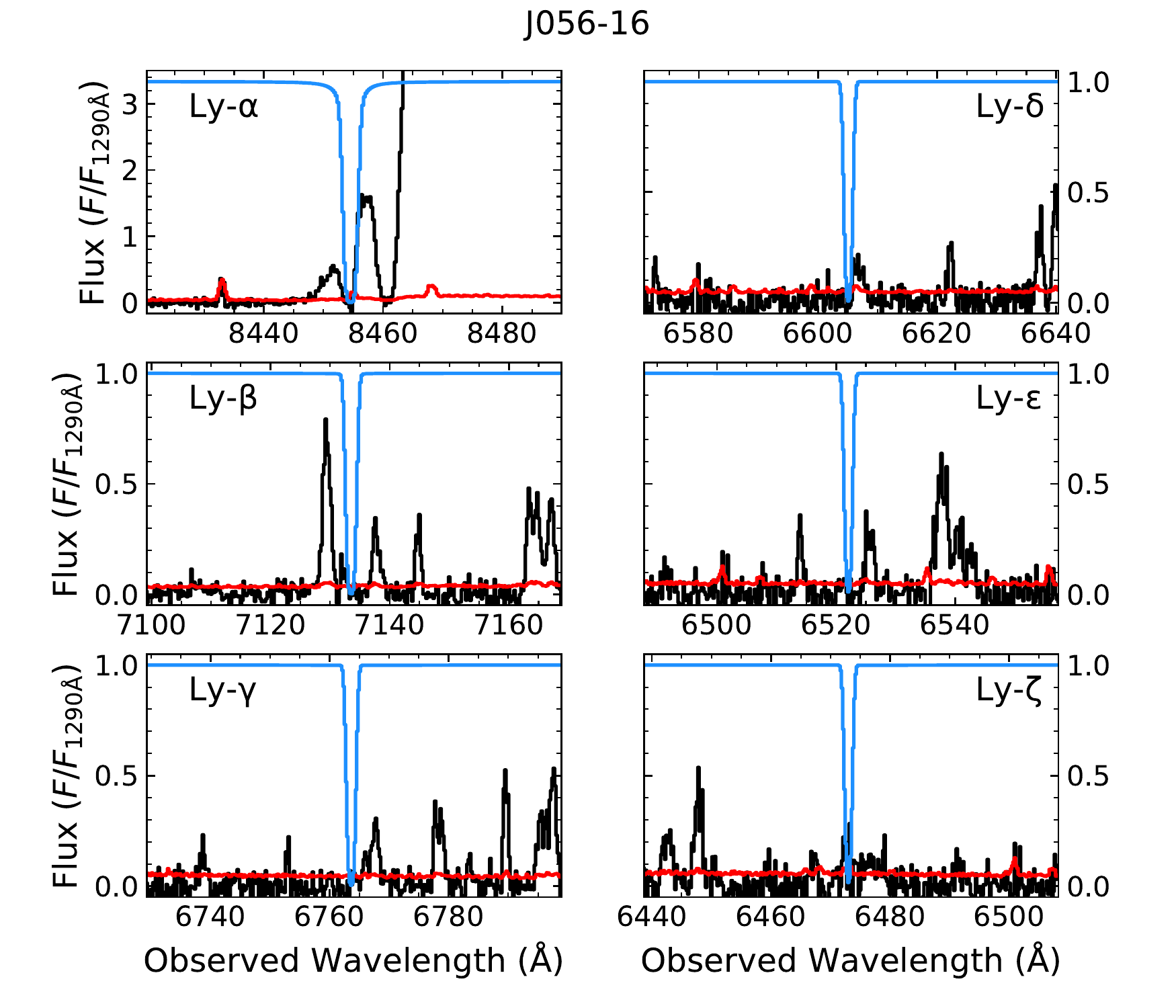}
    \includegraphics[width=0.49\columnwidth]{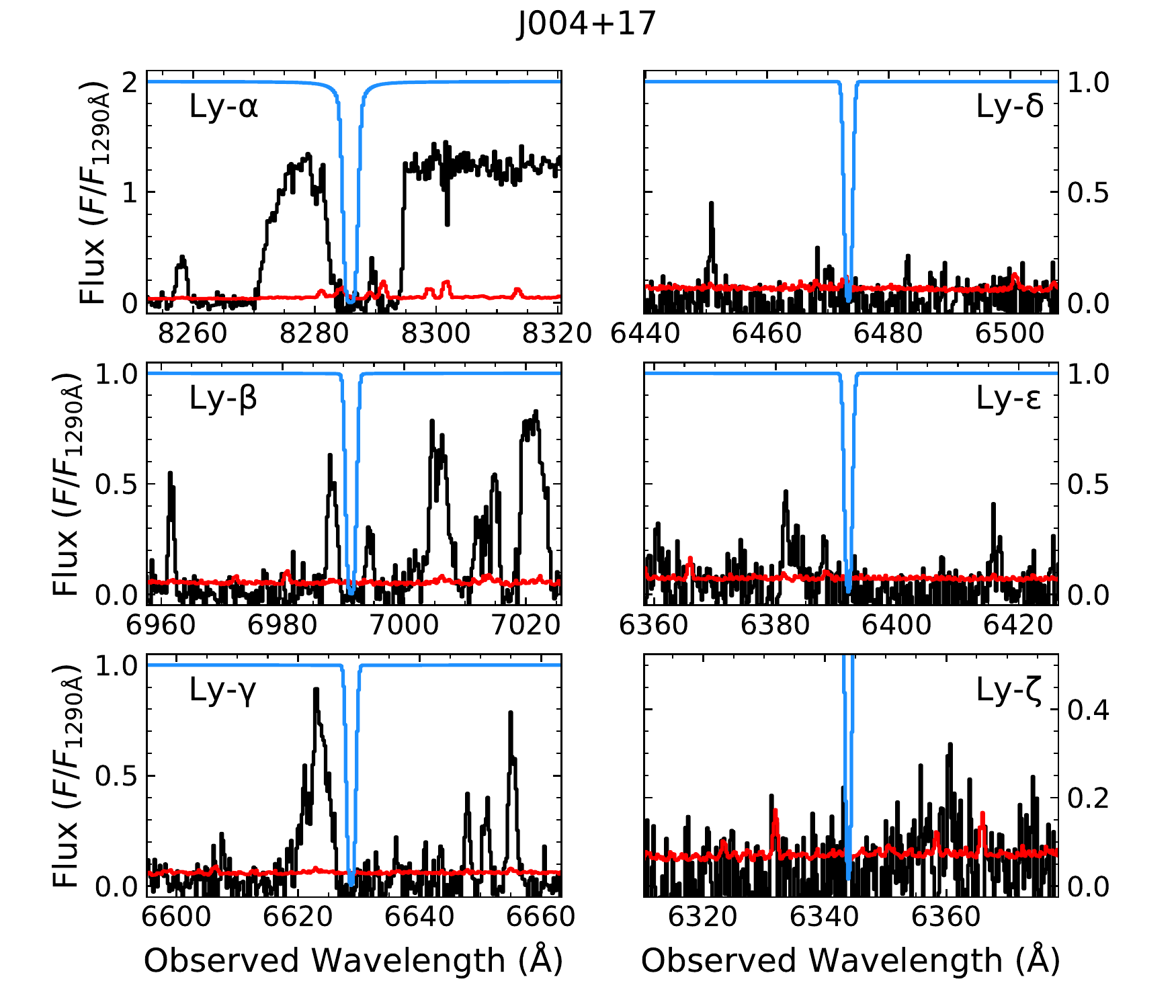}
    \includegraphics[width=0.49\columnwidth]{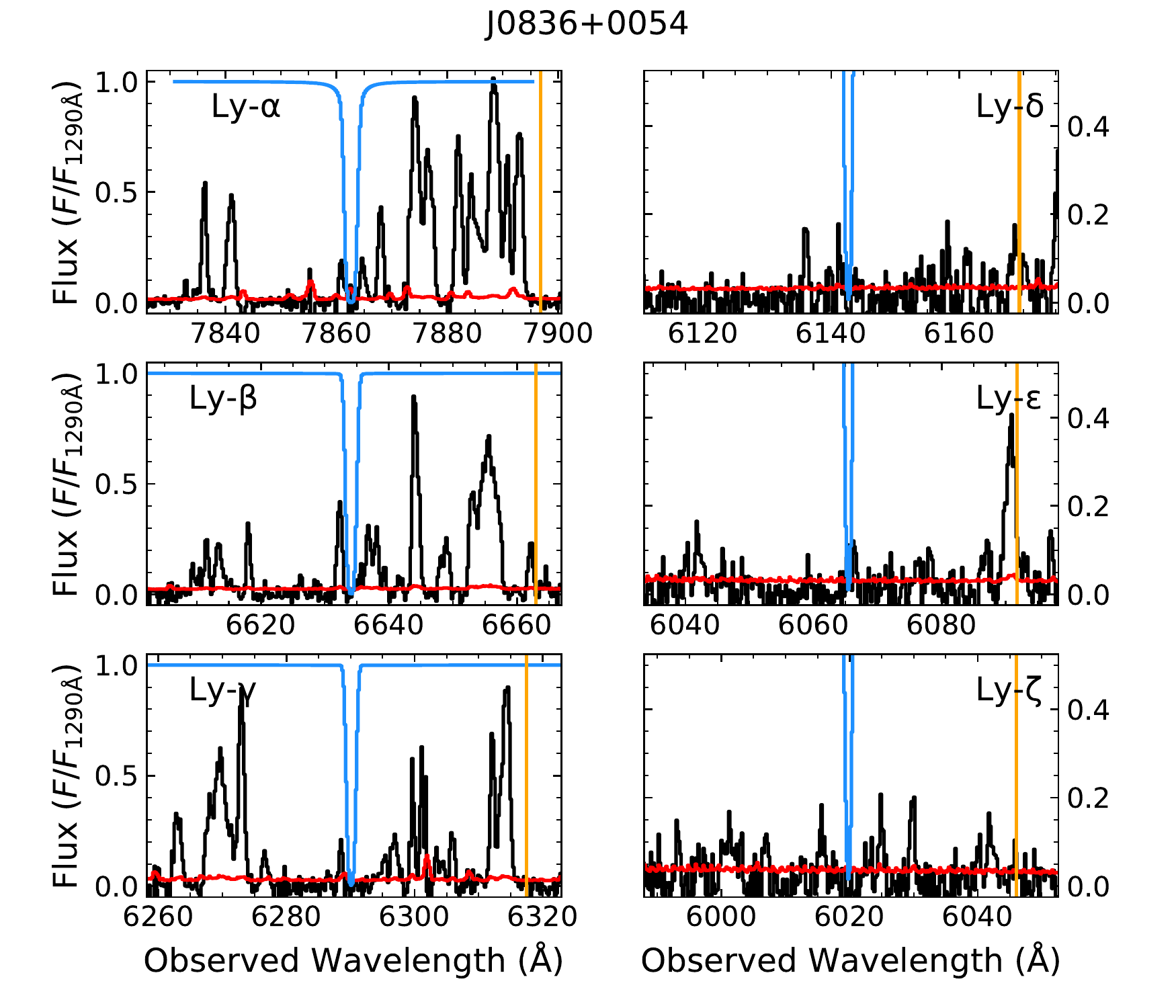}
    \includegraphics[width=0.49\columnwidth]{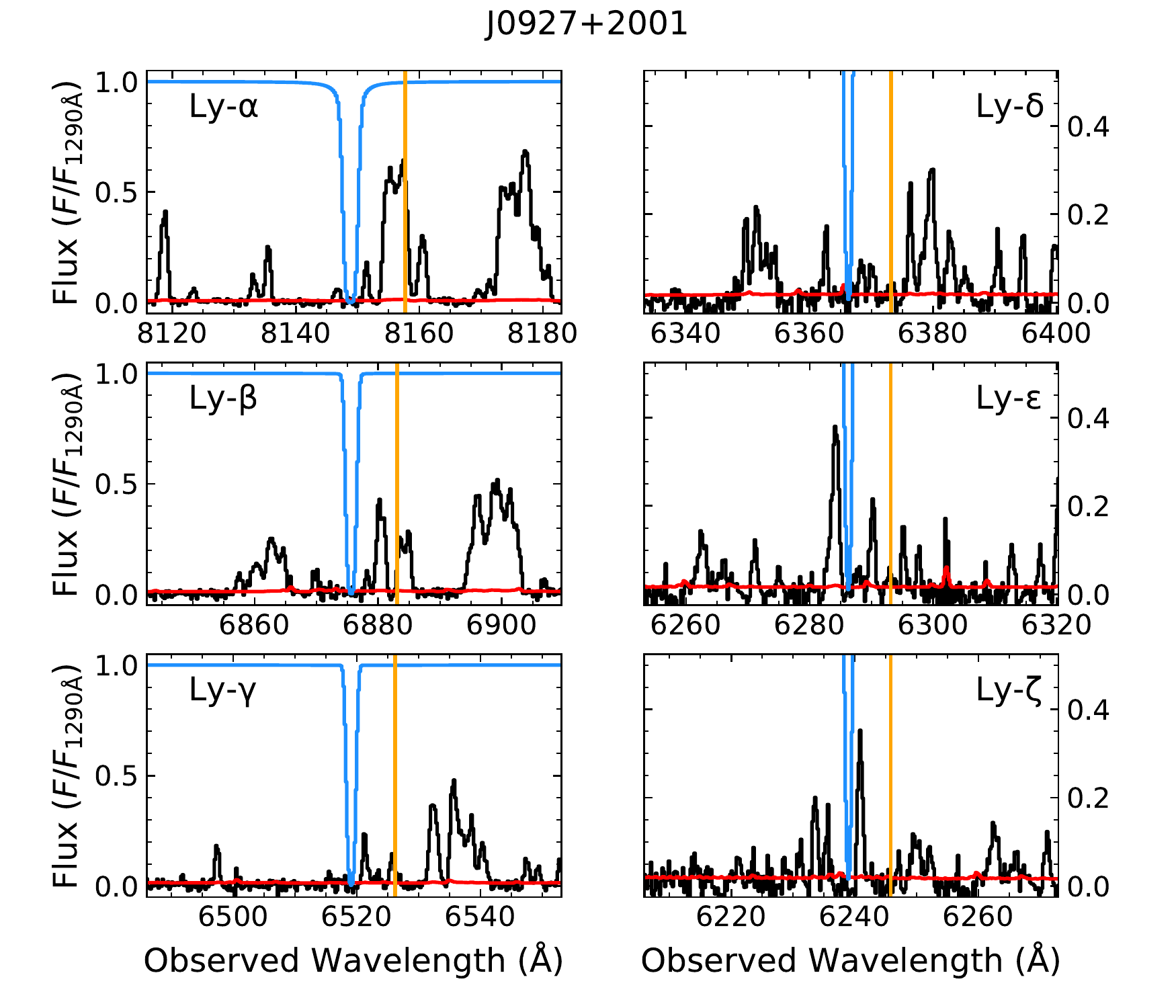}
    \includegraphics[width=0.49\columnwidth]{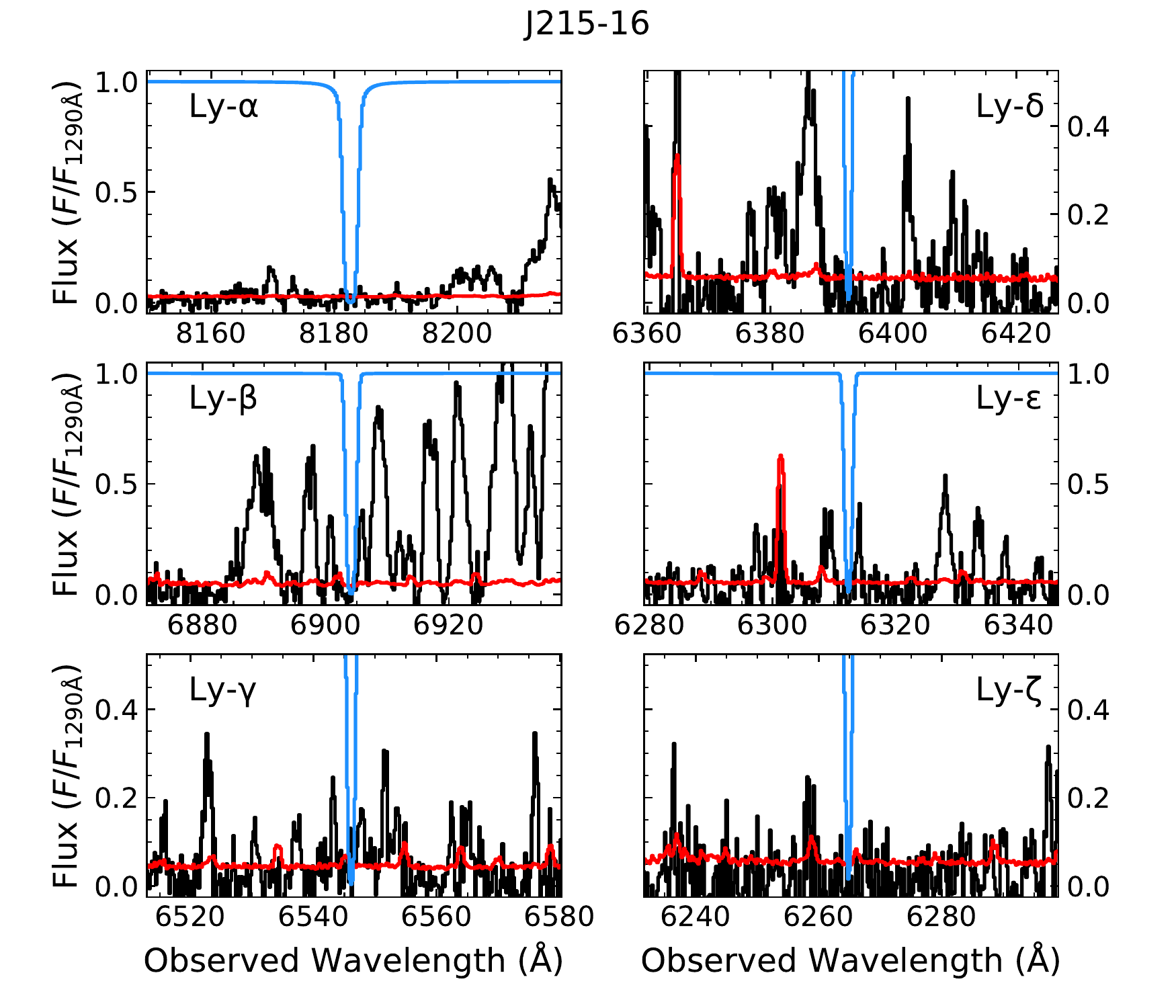}
    \includegraphics[width=0.49\columnwidth]{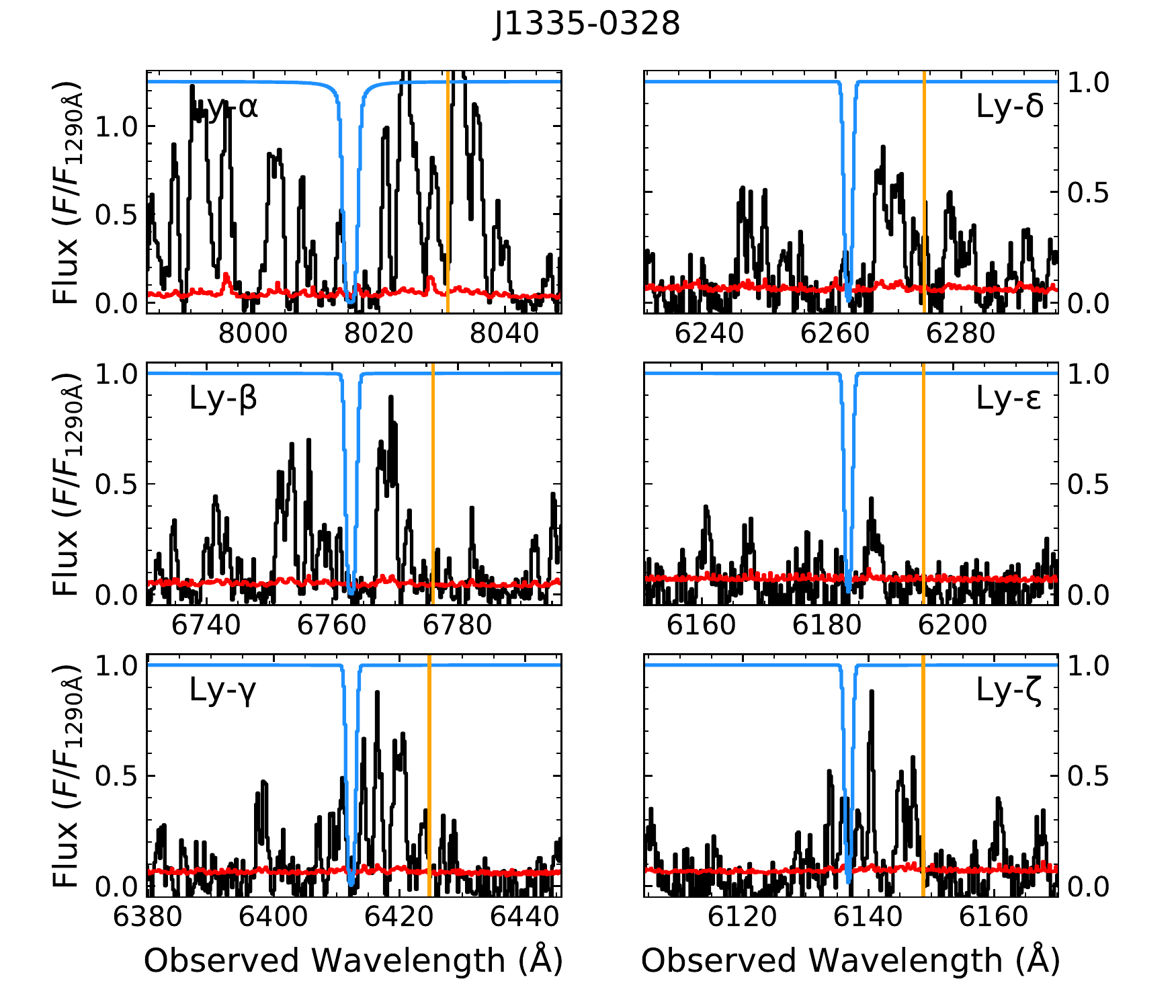}
    \caption{{\textit{continued}}}
\end{figure*}
\begin{figure*}
    \figurenum{4}
    \centering
    \includegraphics[width=0.49\columnwidth]{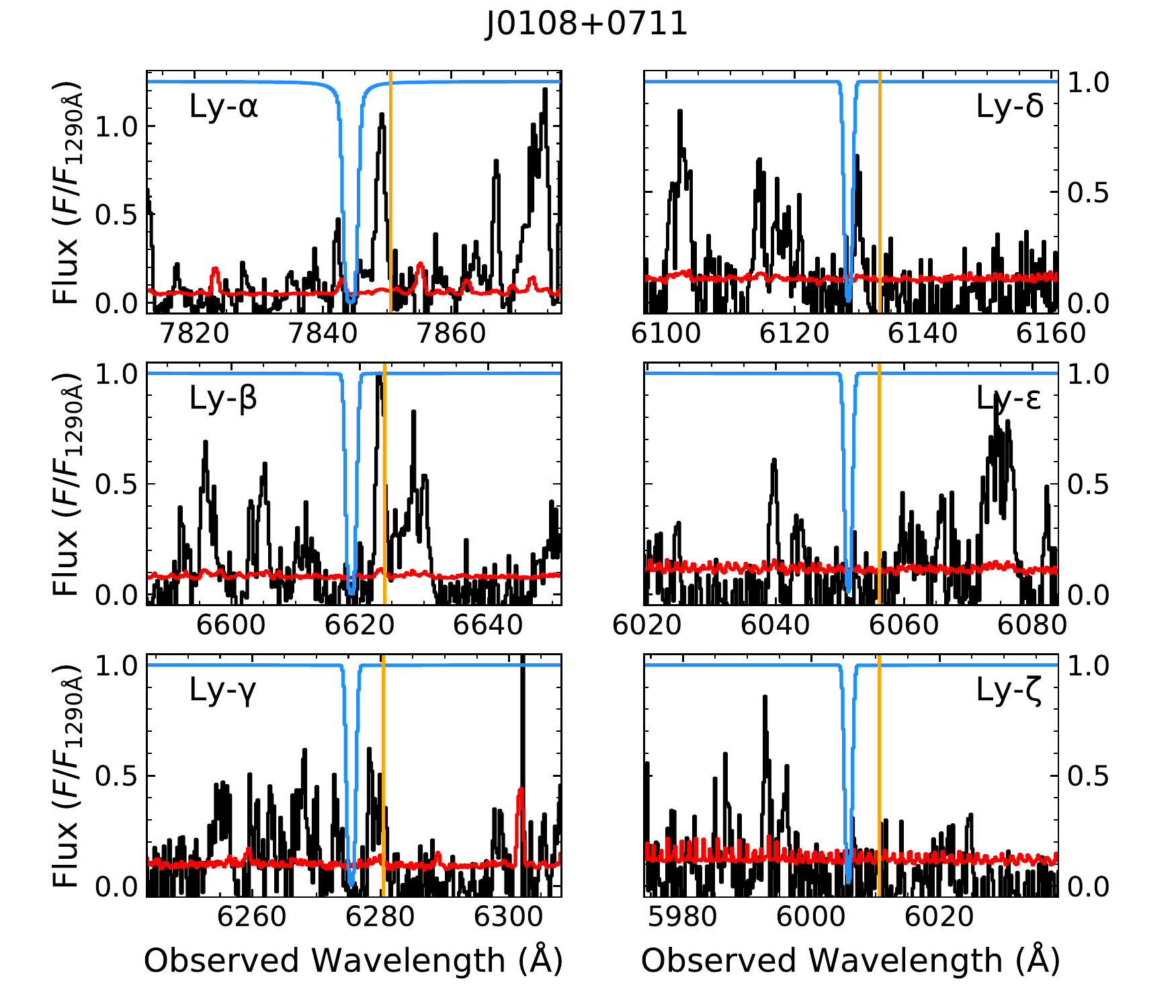}
    \includegraphics[width=0.49\columnwidth]{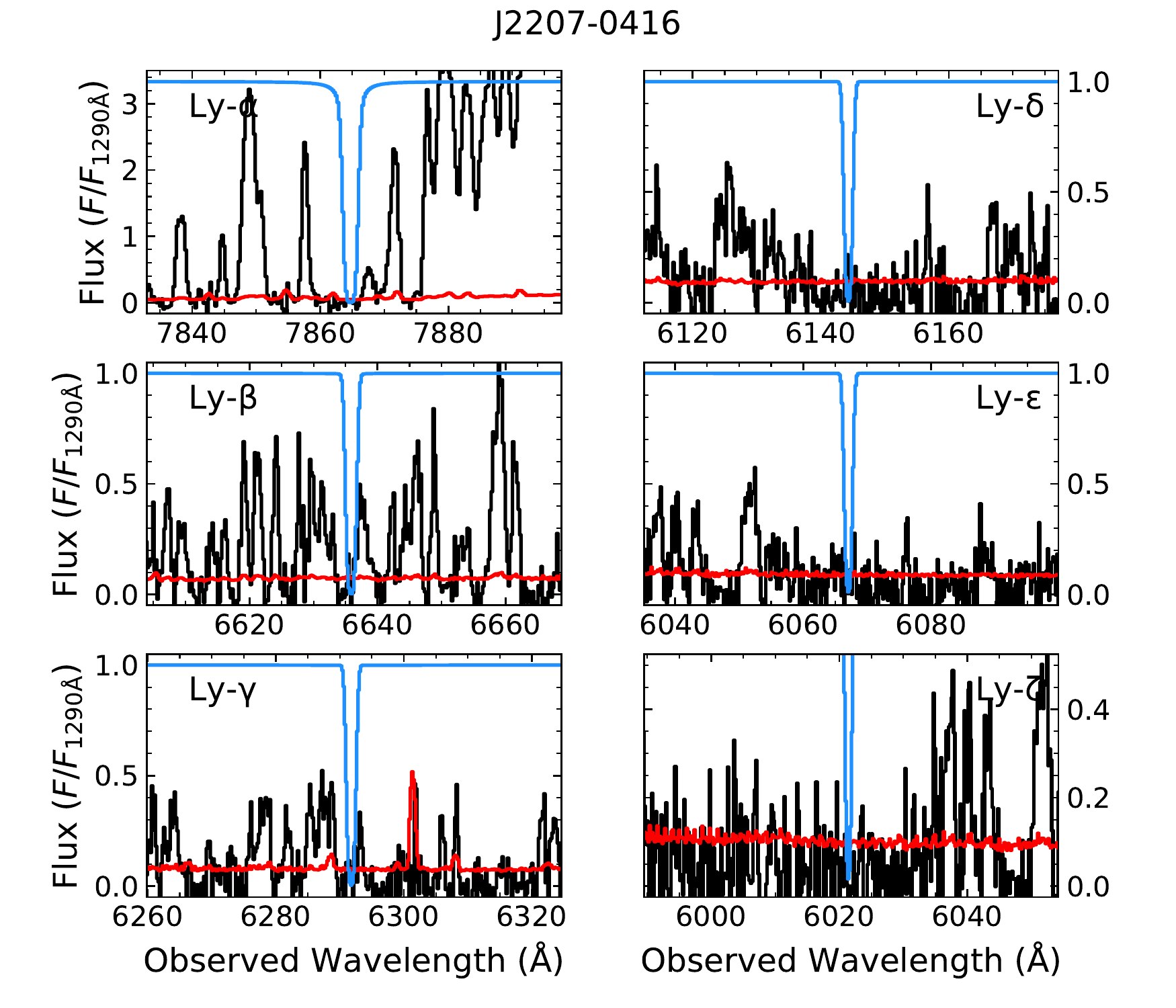}
    \caption{{\textit{continued}}}
    \label{fig:all_transitions}
\end{figure*}

\section{Lyman continuum transmission spikes}\label{sec:app2}

\begin{figure*}
    \figurenum{5}
    \centering
    \includegraphics[width=0.32\columnwidth]{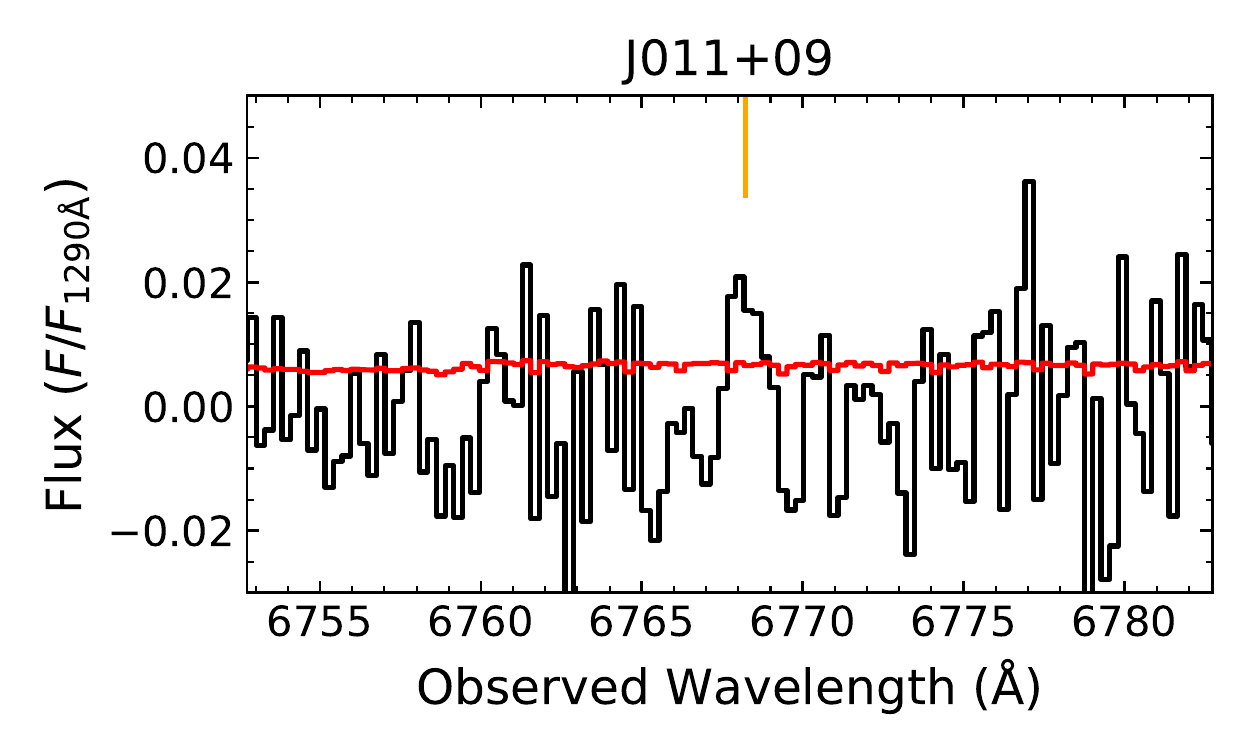}
   \includegraphics[width=0.32\columnwidth]{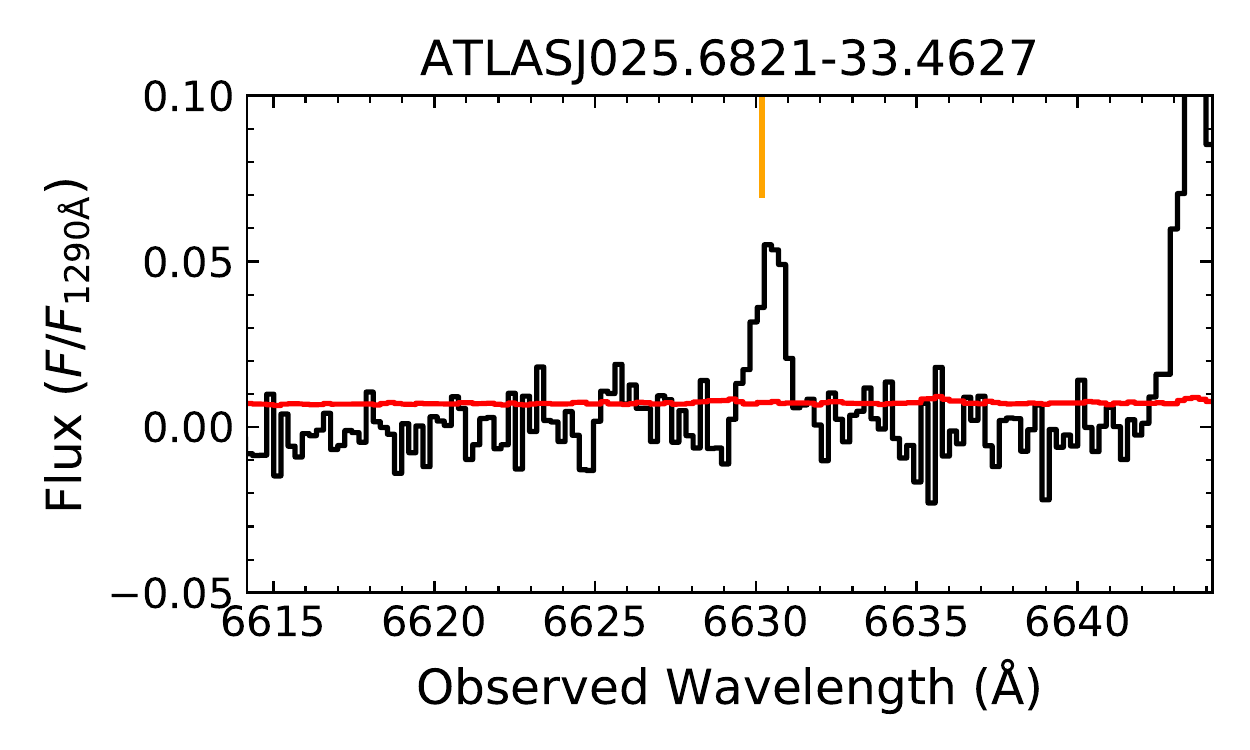}
   \includegraphics[width=0.32\columnwidth]{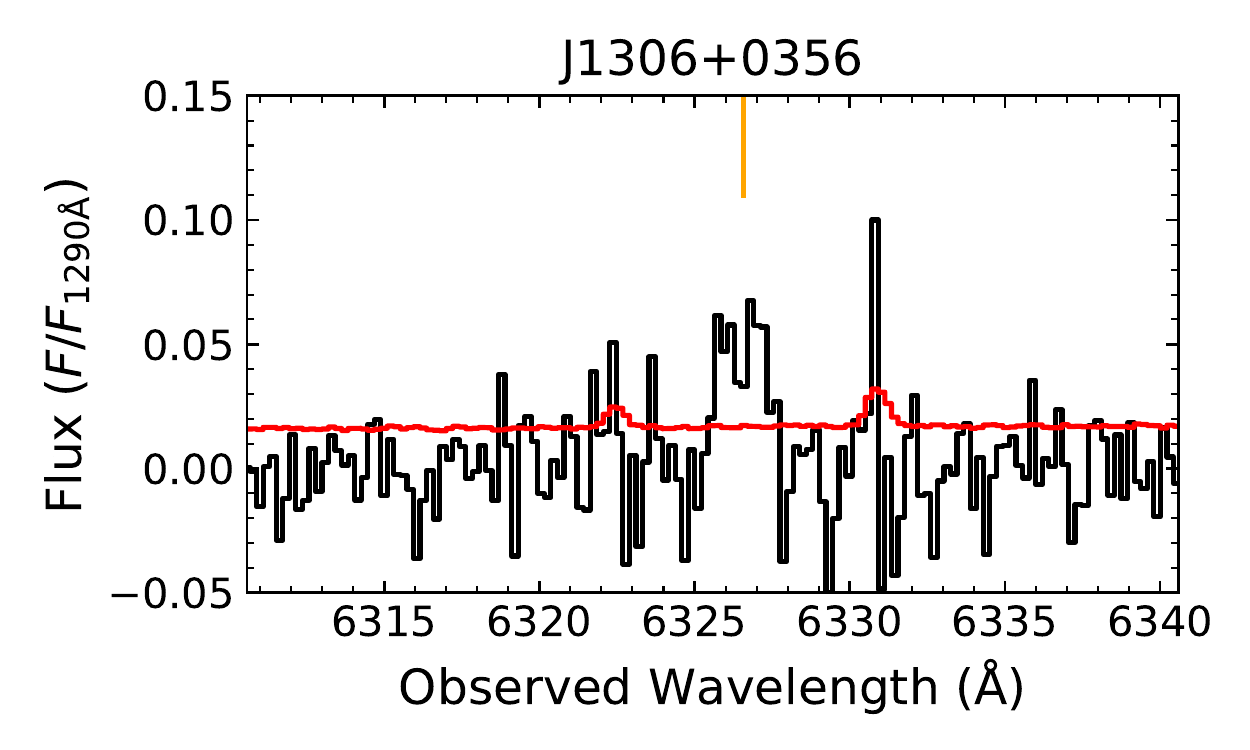}
   \includegraphics[width=0.32\columnwidth]{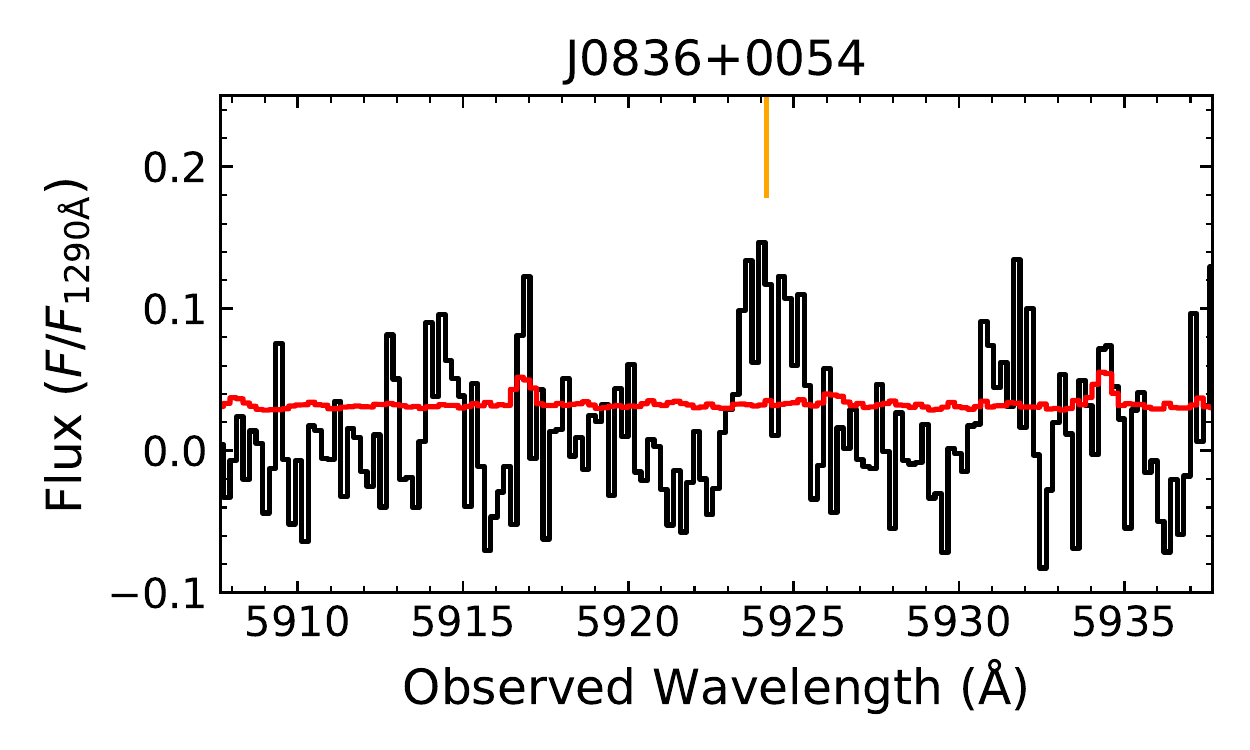}
   \includegraphics[width=0.32\columnwidth]{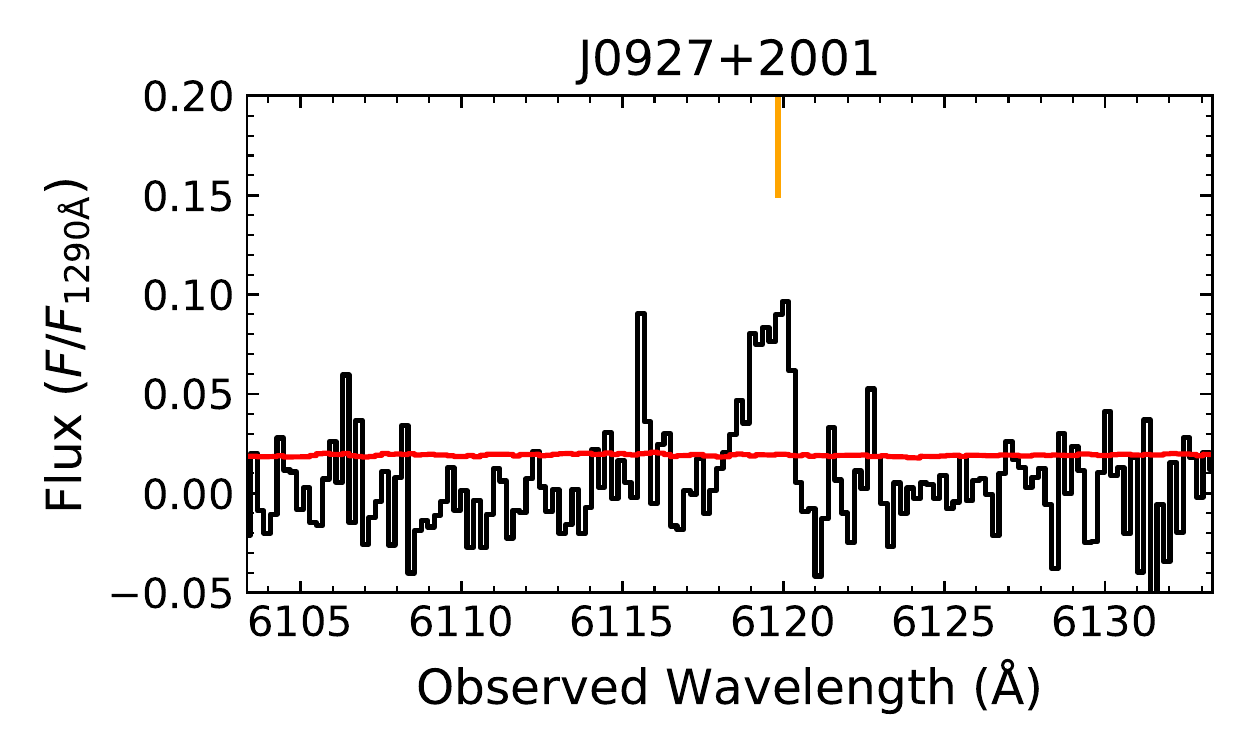}
   \includegraphics[width=0.32\columnwidth]{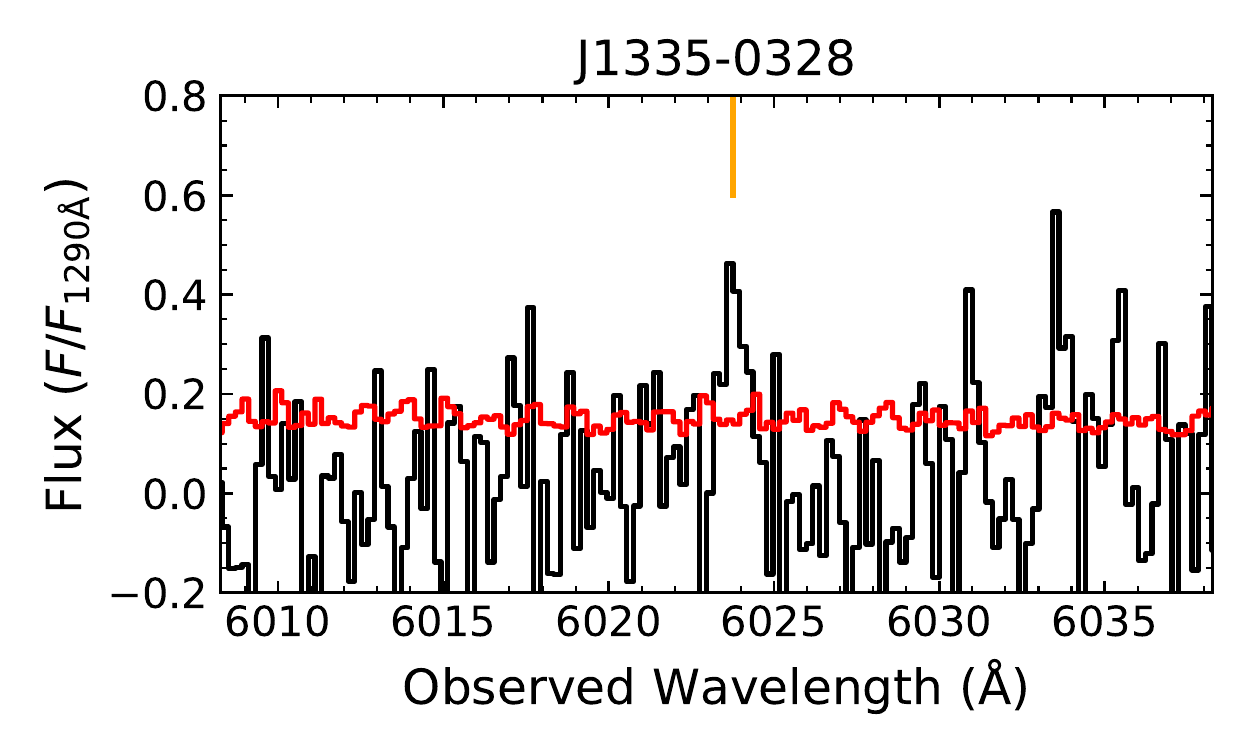}
   \includegraphics[width=0.32\columnwidth]{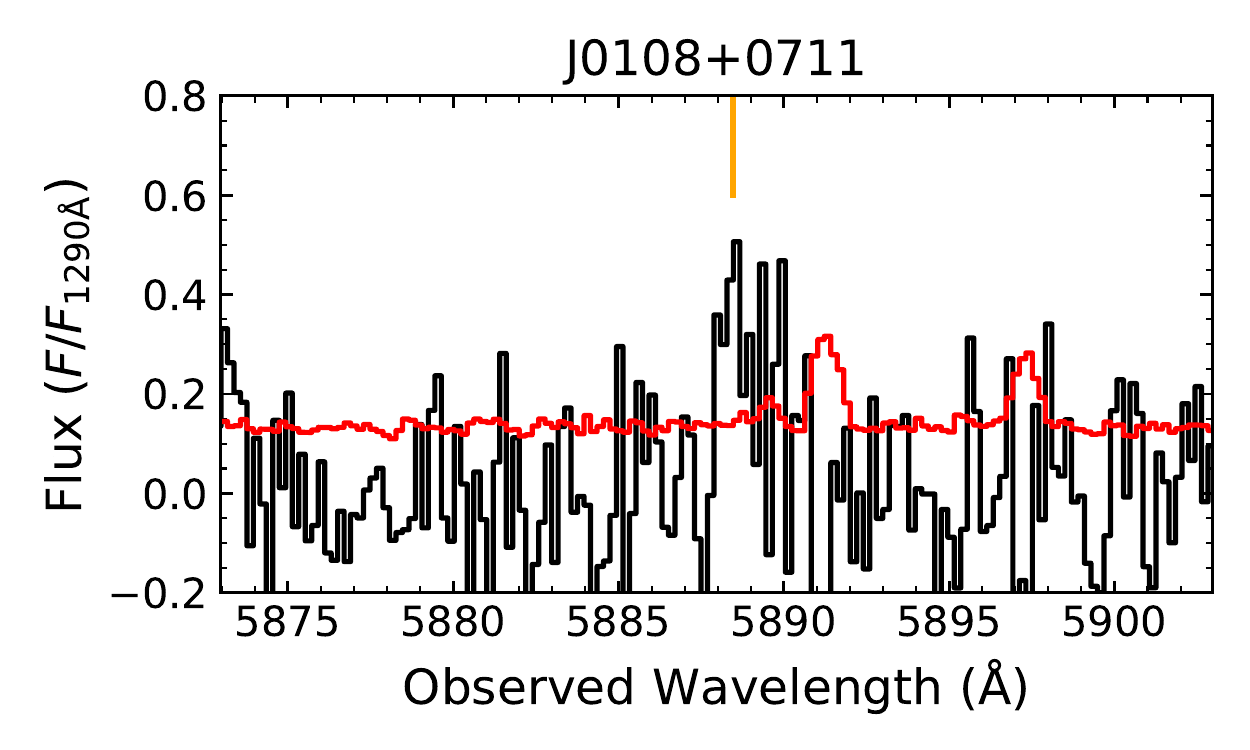}
    \caption{Locations of the Lyman-continuum transmission spikes in the $7$ quasars for which such features helps strengthen the lower limits on LLS distance (see text).}
    \label{fig:spikes}
\end{figure*}

Figure~\ref{fig:spikes} shows the Lyman-continuum transmission detected in $7$ quasars as listed in Table~\ref{tab:data}. The criteria for detection are features detected in $6$ consecutive pixels at $1\sigma$, and a significance of $4\sigma$ for the emission feature overall.

\bibliography{bibliography}
\bibliographystyle{aasjournal}



\end{document}